\documentclass{aa}  

\usepackage[version=3]{mhchem}
\usepackage{graphicx}
\usepackage{txfonts}
\begin{document}

   \title{Millimeter- and submillimeter-wave spectroscopy of thioformamide and interstellar 
   search toward Sgr~B2(N)\thanks{Table A.1 is only available in electronic form at the CDS via anonymous 
   ftp to cdsarc.u-strasbg.fr (130.79.128.5) or via http://cdsweb.u-strasbg.fr/cgi-bin/qcat?J/A+A/}}


        \author{ R.~A. Motiyenko\inst{1}
        \and A. Belloche \inst{2}
        \and R.~T. Garrod \inst{3}
        \and L. Margulès \inst{1}
    \and H.~S.~P.~M{\"u}ller\inst{4}
    \and K.~M. Menten \inst{2}
    \and J.-C. Guillemin\inst{5}
        }

                \institute{Université de Lille, Faculté des Sciences et Technologies, Département Physique, 
                Laboratoire de Physique des Lasers, Atomes et Molécules, UMR CNRS 8523, 
                59655 Villeneuve d'Ascq Cedex, France. \email{roman.motiyenko@univ-lille.fr}
                \and 
                Max-Planck-Institut f\"{u}r Radioastronomie, Auf dem H\"{u}gel 69, 53121 Bonn, Germany
                \and
                Departments of Chemistry and Astronomy, University of Virginia, Charlottesville, VA 22904, USA                  
                \and 
                I. Physikalisches Institut, Universit{\"a}t zu K{\"o}ln, Z{\"u}lpicher Str. 77, 50937 K{\"o}ln, Germany
         \and           
                Univ Rennes, Ecole Nationale Supérieure de Chimie de Rennes, CNRS, ISCR – UMR 6226, 35000 Rennes, France. \\
        }

   \date{Received ; accepted }
        
        \titlerunning{Spectroscopy of thioformamide and the search for it in the ISM}
        
        \authorrunning{R.A. Motiyenko et al.}

  \abstract
   {Thioformamide \ce{NH2CHS} is a sulfur-bearing analog of formamide \ce{NH2CHO}. The latter was detected 
   in the interstellar medium back in the 1970s. Most of the sulfur-containing molecules detected in the interstellar 
   medium are analogs of corresponding oxygen-containing compounds. Therefore, thioformamide is an interesting 
   candidate for a search in the interstellar medium.}
   {A previous study of the rotational spectrum of thioformamide was limited to frequencies below 70~GHz and to 
   transitions with $J \leq 3$. The aim of this study is to provide accurate spectroscopic parameters and 
   rotational transition frequencies for thioformamide to enable astronomical searches for this molecule 
   using radio telescope arrays at millimeter wavelengths.}
   {The rotational spectrum of thioformamide was measured and analyzed in the frequency range 150 to 660~GHz using 
   the Lille spectrometer. We searched for thioformamide toward the high-mass star-forming region Sagittarius (Sgr)~B2(N) 
   using the ReMoCA spectral line survey carried out with the Atacama Large Millimeter/submillimeter Array (ALMA). }
   {Accurate rigid rotor and centrifugal distortion constants were obtained from the analysis of the ground state of 
   parent, $^{34}$S, $^{13}$C, and $^{15}$N singly substituted isotopic species of thioformamide. In addition, for the 
   parent isotopolog, the lowest two excited vibrational states, $\varv_{12}=1$ and $\varv_{9}=1,$ were analyzed using 
   a model that takes Coriolis coupling into account. Thioformamide was not detected toward the hot cores Sgr~B2(N1S) and 
   Sgr~B2(N2). The sensitive upper limits indicate that thioformamide is nearly three orders of magnitude at least less 
   abundant than formamide. This is markedly different from methanethiol, which is only about two orders of magnitude 
   less abundant than methanol in both sources.}
   {The different behavior shown by methanethiol versus thioformamide may be caused by the preferential formation of the latter (on grains) at late times and low temperatures, when CS abundances are depressed. This reduces the thioformamide-to-formamide ratio, because the HCS radical is not as readily available under these conditions.}

   \keywords{astrochemistry -- line: identification -- 
             radio lines: ISM --
             ISM: molecules -- 
             ISM: individual objects: \object{Sagittarius B2(N)}
             }

   \maketitle
%

\section{Introduction}

Several small sulfur compounds have recently been observed in the interstellar medium (ISM): 
\ce{HS2} \citep{fuente2017}, \ce{NS+} \citep{cernicharo2018d}, HCS, and HSC \citep{agundez2018d}. 
To date, 23 sulfur derivatives have been detected or tentatively detected in the ISM, which means that 
the sulfur atom holds the fifth position after hydrogen, carbon, oxygen, and nitrogen atoms; see, for instance 
``Molecules in Space''\footnote{https://cdms.astro.uni-koeln.de/classic/molecules} in the Cologne 
Database for Molecular Spectroscopy. While the first four elements are represented in molecules 
ranging from 2 to 13 atoms with the exception of fullerenes \citep{mcguire2018c}, sulfur is only 
present in small molecules with 2-4 or 6 and possibly 9 atoms with the tentatively 
detected ethanethiol \citep{kolesnikova2014}. For all of the detected sulfur compounds,  
the corresponding oxygen derivatives are observed in the ISM, with the  
exceptions of HSC \citep{agundez2018d} and \ce{C5S} \citep{agundez2014}.

The current question is to know if the larger sulfur derivatives \citep{Herbst09} are difficult to form in the ISM, or have a too short lifetime in interstellar clouds to have sufficiently high abundances allowing their detection. An alternative explanation would be
that spectroscopic studies of such compounds have not been developed enough to enable their detection in the ISM.

We have recently investigated the millimeter spectra of several sulfur compounds with a relatively small 
number of atoms, whose corresponding oxygen derivative has been detected in the ISM and that are among the 
thermodynamically most stable compounds for a given formula: thioacetaldehyde \citep[\ce{CH3C(S)H},][]{margules2020a}, 
propynethial \citep[\ce{HC\bond{3}CCHS},][]{margules2020b}, and S-methyl thioformate \cite[\ce{CH3SC(O)H},][]{jabri2020}. 
We report here the millimeter- and submillimeter-wave spectrum of thioformamide (\ce{NH2CHS}), a six-atom compound. 
Formamide (\ce{NH2CHO}), the corresponding oxygen derivative, has been detected in the ISM as early as 1971 
\citep{rubin1971}, at a time when only CS \citep{penzias1971} and OCS \citep{jefferts1971} had been found as interstellar 
sulfur derivatives. The ground vibrational state of formamide, as well as its lowest excited vibrational states, 
and the ground states of the most abundant isotopic species have been extensively studied \citep{kryvda2009, kutsenko2013}. 
One of these recent studies resulted in the detection of rotational lines of the $\nu_{12}=1$ excited vibrational state 
of formamide toward Orion KL \citep{motiyenko2012}. Nearly 50 years after the first detection of formamide, and after 
the discovery of more than 200 compounds in the ISM, including more than 10\% of sulfur-bearing molecules, thioformamide 
is a relevant candidate for an interstellar detection. Recently detected small molecules containing a sulfur atom 
such as HCS or HSC could be precursors of thioformamide in the ISM by reaction with ammonia, amidogen, or imidogen. 

The microwave spectrum of thioformamide has previously been recorded and analyzed in the frequency range up to 70~GHz 
by \cite{sugisaki1974}. For the ground vibrational state, the analysis included the transitions with $J\leq 3$ and $K_{a} \leq 1$. 
The results of the analysis are therefore not sufficient to extrapolate the spectral predictions of thioformamide over an 
extended range of frequencies and quantum numbers. The results of \citet{sugisaki1974} represent a good starting point 
for our study, in which the rotational spectra were measured in the frequency range from 150 to 660~GHz. 

Sagittarius (Sgr) B2(N) is a protocluster that is forming high-mass 
stars. It is located close to the center of our Galaxy, at a distance of 
8.2~kpc from the Sun \citep{Reid19}. Sgr~B2(N) harbors several hot molecular 
cores \citep[e.g.,][]{Bonfand17,SanchezMonge17}, the main cores are
Sgr~B2(N1) and Sgr~B2(N2). Their high column densities have enabled the 
detection of numerous COMs in their hot inner regions 
\citep[e.g.,][]{Belloche13}. We have recently carried out two sensitive imaging
spectral line surveys of Sgr~B2(N) with the Atacama Large 
Millimeter/submillimeter Array (ALMA) in the 3~mm atmospheric window. These 
surveys, called EMoCA 
\citep[][]{Belloche16} and ReMoCA \citep[][]{Belloche19}, which stand for
exploring molecular complexity with ALMA and reexploring molecular complexity 
with ALMA, respectively, have led in particular to the detection of 
several new complex organic molecules \citep[][]{Belloche14,Belloche17,Belloche19}. Here, we use the latest survey, 
ReMoCA, to search for thioformamide toward Sgr~B2(N).

The article is structured as follows. Section~\ref{s:experiment} 
describes the experiment. The spectroscopic analysis of thioformamide is 
presented in Sect.~\ref{s:spectro}. The results of our search for 
thioformamide in the high-mass star-forming region Sgr~B2(N) are reported in 
Sect.~\ref{s:astro} and are discussed in the context of astrochemical models
in Sect.~\ref{s:discussion}. The conclusions of this work are stated in 
Sect.~\ref{s:conclusions}.

\section{Experiment}
\label{s:experiment}
\subsection{Synthesis}
The synthesis of \cite{londergan1953} has been modified to obtain a formamide-free sample. In a 500~mL 
three-necked flask equipped with a mechanical stirrer and a thermometer, we introduced 300 mL of dry 
tetrahydrofuran and formamide (30~g, 0.667 mol). Phosphorus pentasulfide (\ce{P4S10}, 44.5~g, 0.1~mol) was 
added to the stirred solution in portions of about 5~g during one hour at 30--35$^{\circ}$C. After 6~hours 
of stirring at room temperature, a sticky solid that gradually formed in the reaction mixture was discarded. 
The yellow solution was then transferred under nitrogen in a flask and can be kept for months in the freezer. 
To obtain a pure sample of thioformamide, 30 mL of solution were introduced at room temperature under nitrogen 
in a dry one-necked flask equipped with a stirring bar and a stopcock. The solvent was removed in vacuo (0.1~mbar) 
with the flask immersed in a bath at 30$^{\circ}$C. The flask was then fitted on the spectrometer, immersed in a 
bath heated at 60$^{\circ}$C, and thioformamide was slowly vaporized in the absorption cell of the spectrometer.

\begin{figure*}[htbp]
\begin{center}
\includegraphics[width=\textwidth]{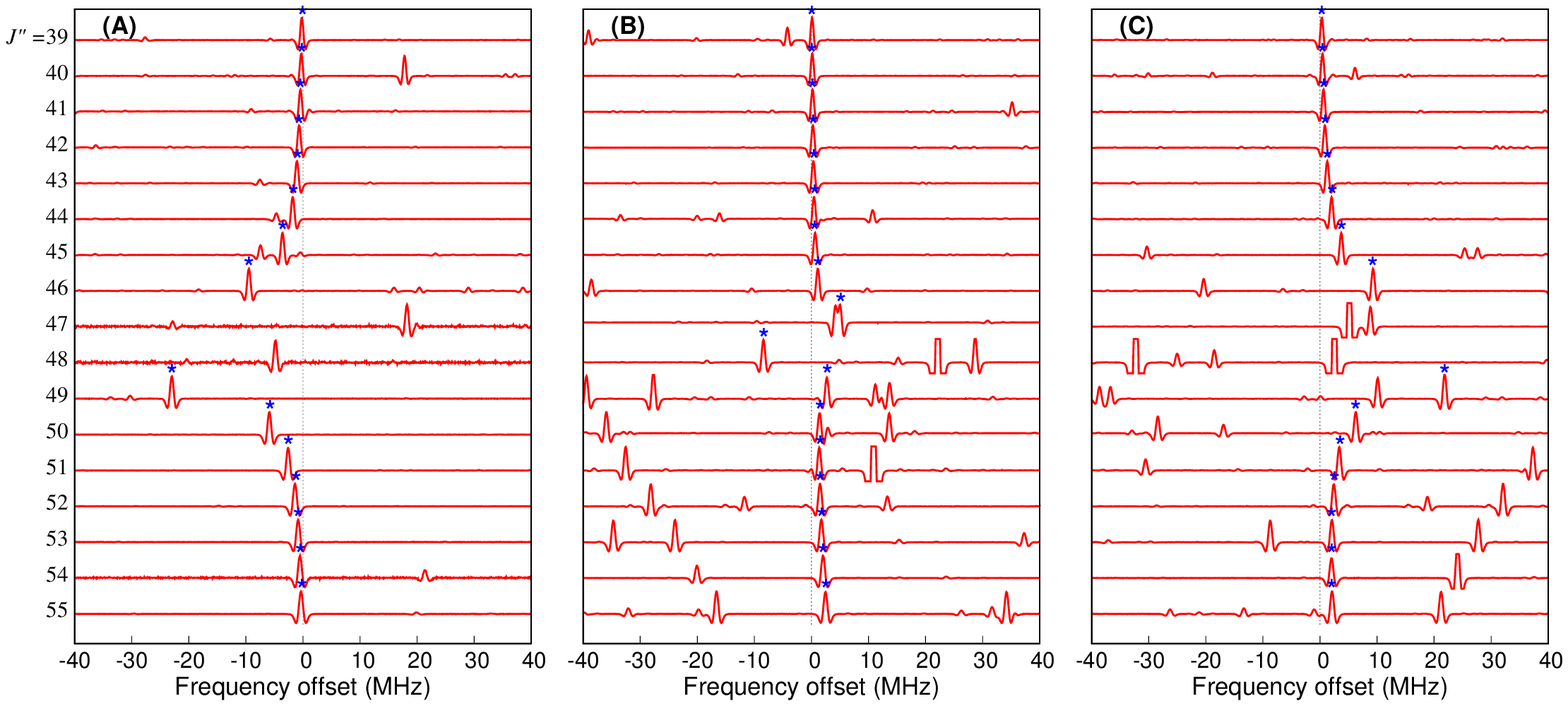}
\caption{Loomis-Wood plots for $J''=39$ to $J'' = 55$ $^{a}R_{0,1}$ series of excited vibrational state 
transitions of thioformamide: (A) $K_{a}=5$, $K_{c}=J''-K_{a}$ of $\nu_{12}=1$ state, (B) $K_{a}=0$ of 
$\nu_{9}=1$ state, and (C) $K_{a}=1$, $K_{c}=J''-K_{a}+1$ of $\nu_{9}=1$ state. Blue stars indicate the lines 
that were effectively assigned to each series. For series (A) and (C), the $J''=47$ and $J''=48$ transitions 
are out of the plot range by $\pm 40$~MHz. }
\label{fig:loomiswood}
\end{center}
\end{figure*}

\subsection{Spectroscopy}

The measurements in the frequency range under investigation were performed using the Lille spectrometer 
\citep{zakharenko2015}, equipped with a fast-scan mode \citep{motiyenko2019}. The frequency ranges 
150$-$330 and 400$-$660~GHz were covered with various active and passive frequency multipliers from 
VDI Inc., and an Agilent synthesizer (12.5$-$18.25~GHz) was used as the reference radiation source. 
The absorption cell was a stainless-steel tube (6 cm diameter, 220 cm long). During measurements, 
the sample was kept at a pressure of about 10 Pa and at room temperature, the line width was determined 
mostly by Doppler broadening. Estimated uncertainties for measured line frequencies are 30~kHz, 50~kHz, 
100~kHz, and 200~kHz  depending on the observed signal-to-noise ratio (S/N) and the frequency range.  

\section{Spectroscopic analysis}
\label{s:spectro}

\subsection{Ground states of parent, $^{34}$S, $^{13}$C, and $^{15}$N isotopic species }

Thioformamide is a prolate asymmetric top close to the symmetric top limit ($\kappa \approx 0.98$). 
Structural parameters obtained from an experimental study \citep{sugisaki1974} and theoretical 
calculations \citep{kowal2006} suggest that the molecule is planar. As was determined by 
\cite{sugisaki1974}, thioformamide has a strong $a$-component of the dipole moment, 
$\mu_{a} = 3.99(2)$~D, and a weak $b$-component $\mu_{b} = 0.13(25)$~D. 

For the initial assignment of the rotational spectrum of thioformamide, we used spectral 
predictions calculated using a set of rigid rotor and quartic centrifugal distortion constants. 
The rigid rotor constants were taken from the previous study \citep{sugisaki1974}. The quartic 
centrifugal distortion constants were obtained from the theoretical calculations of 
harmonic force field preceded by molecular geometry optimization. The theoretical 
calculations were performed using the density functional theory (DFT) employing Becke's 
three-parameter hybrid functional \citep{becke1988}, and the Lee, Yang, and Parr correlation 
functional (B3LYP) \citep{lee1988}. The 6-311++G(3df, 2pd) wave function was employed 
in the B3LYP calculations.

On the basis of the calculated predictions, the assignment of the ground-state transitions 
of the parent isotopic species of thioformamide was straightforward as the frequencies 
of transitions with low $K_{a}$ values were predicted within a few MHz. Several refinements 
of the Hamiltonian parameter data set were needed to accurately fit and predict the transitions 
with higher $K_{a}$ values. In addition to the parent isotopolog, we were able to assign 
the lines of the $^{34}$S, $^{13}$C, and $^{15}$N isotopic species of thioformamide in 
natural abundance, which is correspondingly 4.25\%, 1.1\%, and 0.4\%. As for the ground 
state of the parent species, the initial assignment of isotopologs was based on spectral 
predictions obtained from a set of rigid rotor constants determined by \citet{sugisaki1974} 
and centrifugal distortion constants of the parent species determined in this study. 
We note that the assignment of the isotopic species was greatly facilitated by 
the results of \citet{sugisaki1974}. Owing to strong spectral congestion of weak lines 
caused in particular by sample impurities, it would be practically impossible to 
distinguish characteristic spectral patterns of isotopologs. Therefore the 
assignment was only possible by direct comparison of the experimental spectra 
with calculated predictions. 

All the assigned transitions were fit to a Watson $A$-reduction Hamiltonian 
in $I^{r}$ coordinate representation. In addition, we also tested $S$-reduction 
Hamiltonian, as the asymmetry parameter of thioformamide is quite close to the 
symmetric top limit. We applied both reductions in the least-squares fit of the 
assigned ground-state transitions of the parent isotopolog, and obtained the fits
of similar quality in terms of root-mean-square and weighted root-mean-square 
deviations, and number of lines that were fit. However, in the case of the $S$-reduction 
Hamiltonian, the least-squares fit required one more parameter than the $A$-reduction. 
This allowed us to suggest that for thioformamide, $A$-reduction Hamiltonian may be 
preferred to $S$-reduction. The complete list of measured rotational transitions in 
the ground state of parent isotopic species of thioformamide is presented in Table~\ref{tab:rottrans} 
available at the CDS. Owing to its significant size, we show only a part of 
Table~\ref{tab:rottrans} as an example here.

The results of the fits including $S$- and $A$-reduction tests are given in 
Table \ref{tab:rot}. All the data sets were fit within  experimental accuracy. 
For each isotopic species we determined the full set of fourth$^{\rm }$ order 
centrifugal distortion constants, as well as several sixth-$^{\rm }$ and eighth-order$^{\rm }$ 
(in the case of the parent and $^{34}$S species) constants. 
For the ground state of the parent isotopic species, we were able to assign 
about 30 weak $b$-type transitions. To correctly reproduce the intensities of 
the $b$-type transitions with respect to other assigned neighboring lines, 
we found that the $\mu_{b}$ value should be scaled to about 0.2~D, which is higher 
by a factor of 1.5 than the $\mu_{b}$ value determined by \citet{sugisaki1974}. 
We are not able to provide a more accurate determination of the $\mu_{b}$ value and 
its uncertainty because we were unable to perform a comparison
with the intensities of $a$-type transitions of the ground state of parent 
isotopic species here. The assigned $a$- and $b$-type transitions are significantly 
spaced on the frequency scale, whereas baseline and radiation source intensity 
variations make this comparison very inaccurate.
Moreover, we were unable to resolve the nuclear quadrupole 
hyperfine structure owing to the \ce{^{14}N} atom in the Doppler-limited spectra of thioformamide.

\begin{table*}
\centering
 \caption{Ground-state rotational constants of parent, $^{34}$S, $^{13}$C, and $^{15}$N isotopic species of thioformamide}  
 \label{tab:rot}        
\begin{tabular}{lr|lrrrr}
\hline\hline
\multicolumn{2}{c|}{S-reduction} & \multicolumn{5}{c}{A-reduction} \\
Parameters    &     \multicolumn{1}{c|}{Parent} & Parameters    &     \multicolumn{1}{c}{Parent} &   \multicolumn{1}{c}{\ce{H2NCH^{34}S}}  &    \multicolumn{1}{c}{\ce{H2N^{13}CHS}}  &  \multicolumn{1}{c}{\ce{H{_2}^{15}NCHS}}    \\
\hline                                                                                                
$A$ (MHz)           & 61754.8040(80)\tablefootmark{a}     & $A$ (MHz) & 61754.8024(81)  &  61632.104(56)       &  60002.038(75)        &  61319.746(96)          \\
$B$ (MHz)           &  6101.76611(14)    & $B$ (MHz)             &  6101.80024(14)     &   5952.38931(27)     &    6085.09318(38)     &    5919.64877(56)       \\
$C$ (MHz)           &  5549.86284(13)    & $C$ (MHz)             &  5549.82889(13)     &   5424.98705(26)     &    5521.38188(34)     &    5395.37815(56)       \\
$D_J$ (kHz)         &     2.559762(22)    & $\Delta_J$ (kHz)      &    2.601528(25)     &      2.483353(41)    &      2.561801(64)     &      2.470969(87)       \\
$D_{JK}$ (kHz)      &   -48.89300(51)     & $\Delta_{JK}$ (kHz)   &  -49.14361(51)      &    -48.1177(11)      &    -47.6244(13)       &    -48.9717(24)         \\
$D_K$ (kHz)         &  1508.12(41)        & $\Delta_K$ (kHz)      & 1508.26(42)         &   1522.0(31)         &   1490.7(37)          &   1513.4(57)            \\
$d_1$  (kHz)        &    -0.379115(39)    & $\delta_J$  (kHz)     &    0.379174(36)     &      0.354556(60)    &      0.384478(83)     &      0.35373(14)        \\
$d_2$  (kHz)        &    -0.020822(17)    & $\delta_K$  (kHz)     &   16.9294(35)       &     16.362(10)       &     16.757(13)        &     15.975(21)          \\
$H_J$ (Hz)          &     0.0018292(39)  & $\Phi_J$ (Hz)         &     0.0019788(40)   &      0.0018384(61)   &       0.001885(10)    &       0.001848(13)      \\
$H_{JK}$ (Hz)       &    -0.02017(19)    & $\Phi_{JK}$ (Hz)      &    -0.02012(19)     &     -0.01759(10)     &      -0.01902(22)     &      -0.01806(41)       \\
$H_{KJ}$ (Hz)       &    -5.4240(16)     & $\Phi_{KJ}$ (Hz)      &    -5.4267(16)      &     -5.3629(82)      &      -5.2389(47)      &      -5.241(13)         \\
$H_K$ (Hz)          &   144.5(62)        & $\Phi_K$ (Hz)         &   143.4(63)         &      0.0            &       0.0    &       0.0      \\
$h_1$ (Hz)          &  0.0006741(67)     & $\phi_J$ (Hz)         &     0.0006843(61)   &      0.0006279(98)  &      0.000687(14)    &     0.000643(24)       \\
$h_2$ (Hz)          &  0.0000590(45)     & $L_{JK}$ (mHz)        &    -0.01139(23)     &      0.0            &      0.0            &      0.0                \\ 
$L_{JK}$ (mHz)      & -0.01315(23)       & $L_{KKJ}$ (mHz)       &     0.5306(15)      &      0.499(16)      &      0.0            &      0.0                \\
$L_{KKJ}$ (mHz)     &  0.5352(15)        &                                              &                                                 &                                               &                                       &                                                       \\
\hline                                                                                               
 N\tablefootmark{b}                                  &  1066 &  &    1066     &      773     &      716           &    261        \\
 $J_{\mathrm{max}}$, $K_{a, \mathrm{max}}$  & 59, 27 &   & 59, 27     &      59, 19     &      59, 17           &   58, 15       \\
 $\sigma$ (MHz)\tablefootmark{c}            & 0.028 &   &  0.028   &      0.028    &       0.039         &   0.44          \\
 $\sigma_w$\tablefootmark{d}                &  0.57 &   &   0.57   &      0.56     &        0.76         &     0.86       \\   
\hline                                                                                         
\end{tabular}
\tablefoot{
 \tablefoottext{a}{Number in parentheses are one time the standard deviation.}
 \tablefoottext{b}{Number of distinct frequency lines in fit.}
 \tablefoottext{c}{Standard deviation of the fit.}
 \tablefoottext{d}{Weighted deviation of the fit.}
 }
 
 \end{table*}

\subsection{Excited vibrational states $\varv_{12}=1$, and $\varv_{9}=1$ of the parent isotopic species}

For the parent isotopic species, the rotational transitions
of the lowest excited vibrational states formed quite easily distinguishable patterns
of satellite lines in the observed spectrum. The two lowest vibrational modes of thioformamide are 
\ce{NH2} out-of-plane wagging $\nu_{12}$, and in-plane bending 
of \ce{NCS} moiety $\nu_{9}$. The vibrational frequencies of these modes 
are 393 and 457~cm$^{-1}$, respectively, as determined from the analysis of 
relative intensities by \citet{sugisaki1974}. The analysis of the excited states 
started with the assignment of $K_{a}=0$ series of lines of $\varv_{12}=1$ state 
and continued to higher $K_{a}$ values. The series of $K_{a}=6$ lines was found to be
perturbed as some of the lines exhibited quite large and irregular deviations from 
predicted frequencies. For the $\varv_{9}=1$ we found that even $K_{a}=0$ and $K_{a}=1$ 
series of lines were perturbed in a similar way. 

The perturbations are illustrated on Fig.~\ref{fig:loomiswood}, where 
Loomis-Wood type diagrams are shown for the $^{a}R_{0,1}$ series of 
transitions with $J''$ in the range from 39 to 55, and where $J''$ is the 
upper level quantum number. Each diagram is represented by a superposition 
of experimental spectra centered on the corresponding series transition frequency. 
The transition frequencies were calculated on the basis of the results of a 
single-state fit that does not take any vibrational state coupling into account. 
The analysis of Fig.~\ref{fig:loomiswood} shows that the lines of 
series (A) and (C) that represent $K_{a}=5$, $K_{c}=J''-K_{a}$ transitions 
of $\varv_{12}=1$ state and $K_{c}=J''-K_{a}+1$ of $\varv_{9}=1$ state, 
respectively, significantly deviate from calculated values for $J''$ 
in the range 45 to 51. For the two series, the deviations from calculated 
frequencies are approximately the same, but with opposite sign. In addition, 
relatively smaller deviations are observed for $J''=47$ to 49 of series (B), which 
represents $K_{a}=0$ transitions of the $\varv_{9}=1$ state. This behavior is a 
typical example of local resonances coupling $\varv_{12}=1$ and $v_{9}=1$ states. 
In this case, the coupling Hamiltonian may be presented in the following block-diagonal form:
\begin{equation}
\label{eq:hcoup}
H = \left( \begin{array}{cc}
H_{\mathrm{rot}}^{(12)} & H_{\mathrm{c}} \\
H_{\mathrm{c}} & H_{\mathrm{rot}}^{(9)}+\Delta E \\
\end{array}
\right),
\end{equation} where $H_{\mathrm{rot}}^{(12)}$ and $H_{\mathrm{rot}}^{(9)}$ 
are the standard rotational Watson $A$-reduction Hamiltonians, $\Delta E$ 
is the energy difference between two coupled states, and $H_{\mathrm{c}}$ 
is the off-diagonal interaction term.

When molecular planarity is assumed, the equilibrium configuration of thioformamide 
is described by the $C_s$ point group. The vibrational modes $\nu_{12}$ 
and $\nu_{9}$ belong to $A''$ and $A'$ irreducible representations 
of the group, respectively. As a result, Coriolis interaction between the two modes is 
allowed along the $a$ and $b$ axes, and the Hamiltonian $H_{\mathrm{c}}$ 
we used is expressed as
\begin{equation}
\begin{array}{l}
H_{\mathrm{c}} = i(G_a+G_a^JP^2+G_a^KP_z^2+...)P_z + \\
  i(G_b+G_b^JP^2+G_b^KP_z^2+G_b^{JK}P^2P_z^2...)P_x
\end{array},
\end{equation} where $G_a$ and $G_b$ are the Coriolis coupling constants, 
and all other parameters are their respective centrifugal distortion corrections. 

In search of an initial global solution of the two-state coupling problem 
using the Hamiltonian in Eq.~\ref{eq:hcoup}, we used a method that was 
previously successfully applied for a similar case \citep{motiyenko2018}. 
Because the  $\Delta E$, and $G_a$ and $G_b$ parameters are usually highly 
correlated, and given the absence of sufficiently accurate initial values 
of these parameters, we fixed $\Delta E$ to a series of reasonable values. 
The range of $\Delta E$ values was estimated from the energy difference 
between $\varv_{12}=1$ and $\varv_{9}=1$ states determined by \citet{sugisaki1974}. 
Taking the uncertainties of the vibrational frequencies of the $\nu_{12}$ 
and $\nu_{9}$ modes into account, we generated a set of $\Delta E$ values 
that varied from 35~cm$^{-1}$ to 114~cm$^{-1}$ with a step of about 0.03~cm$^{-1}$. 
For each fixed $\Delta E$ from the set we performed a least-squares fit using 
the Hamiltonian in Eq.~\ref{eq:hcoup} in which $G_a$ and $G_b$ were varied. 
A global solution corresponding to the minimum root-mean-square deviation of 
the fit was found for $\Delta E \approx$~60~cm$^{-1}$. Then, $\Delta E$ was 
allowed to vary in the fit along with $G_a$ and $G_b$ starting from its 
approximate value found at the previous step. In this manner, we were 
able to fit the perturbed transitions shown 
in Fig.~\ref{fig:loomiswood} within experimental accuracy and to accurately predict the most perturbed 
transitions that were out of the range of the Loomis-Wood plot. The following 
assignment process was straightforward and similar to the assignment of the 
ground state with several cycles of refinement of the Hamiltonian parameters. 
The results of the global fit are presented in Table~\ref{tab:v912}. In terms 
of signs and orders of magnitude, the determined centrifugal distortion constants 
of the two excited vibrational states agree with the corresponding 
ground state parameters. The final value of energy difference between $\varv_{12}=1$ 
and $\varv_{9}=1$ states, 59.991696(57)~cm$^{-1}$, also agrees with the 
energy difference of about 64~cm$^{-1}$ from the analysis of relative intensities by \citet{sugisaki1974}.

\begin{table}
\centering
 \caption{Rotational constants of $\varv_{12}=1$, and $\varv_{9}=1$ excited vibrational states of the parent isotopolog of thioformamide.}  
 \label{tab:v912}       
\begin{tabular}{lrrrr}
\hline\hline
 Parameters              &   $\varv_{12} = 1$   &  $\varv_{9} = 1$      \\
\hline                                                                        
 $A$ (MHz)               & 61118.626(83)\tablefootmark{a}      &  61890.776(81)       \\
 $B$ (MHz)               &  6098.12788(52)     &   6101.04527(57)     \\
 $C$ (MHz)               &  5552.46610(25)     &   5544.50067(27)     \\
 $\Delta_J$ (kHz)        &     2.608050(46)    &      2.582672(44)    \\
 $\Delta_{JK}$ (kHz)     &   -48.371(17)       &    -47.945(17)       \\
 $\Delta_K$ (kHz)        &  1433.5(12)         &   1561.6(16)         \\
 $\delta_J$  (kHz)       &     0.377740(70)    &      0.377563(72)    \\
 $\delta_K$  (kHz)       &    16.3958(97)      &     18.7564(88)      \\
 $\Phi_J$ (Hz)           &     0.0019722(68)  &      0.0019123(65)   \\
 $\Phi_{JK}$ (Hz)        &    -0.02115(16)    &     -0.01684(19)     \\
 $\Phi_{KJ}$ (Hz)        &    -4.9201(99)     &     -5.961(12)       \\
 $\phi_J$ (Hz)           &    0.000670(14)  &     0.000648(13)   \\
 $L_{KKJ}$ (mHz)         &    0.423(19)     &     0.450(25)      \\
 $\Delta E$ (MHz) / (cm$^{-1}$)       & \multicolumn{2}{c}{1798505.8(17) / 59.991696(57)  }         \\
 $G_a$ (MHz)             & \multicolumn{2}{c}{25010.4(24)    }    \\
 $G_a^J$ (MHz)           & \multicolumn{2}{c}{   0.00665(59)}    \\
 $G_b$ (MHz)             & \multicolumn{2}{c}{-829.39(51)     }    \\
 $G_b^J$ (MHz)           & \multicolumn{2}{c}{  0.003478(42)}    \\
 \hline                                                                                                       
 N\tablefootmark{b}                      &    690       &      686             \\
 $\sigma$ (MHz)\tablefootmark{c} &    0.028     &       0.027           \\
 $\sigma_w$\tablefootmark{d}     &    0.68      &       0.64            \\   
\hline                                                                                                        
\end{tabular}
\tablefoot{
 \tablefoottext{a}{Number in parentheses are one time the standard deviation.}
 \tablefoottext{b}{Number of distinct frequency lines in the fit.}
 \tablefoottext{c}{Standard deviation of the fit.}
 \tablefoottext{d}{Weighted deviation of the fit.}
 }                                    
 
 \end{table}

\subsection{Rotational spectrum predictions}
On the basis of the parameter set presented in Table~\ref{tab:rot}, 
we calculated spectrum predictions of the ground-state rotational transitions 
of the parent isotopic species of thioformamide. The predictions calculated 
at $T = 300$~K are given in Table~\ref{tab:pred}. Owing to its significant size, 
the complete version of Table~\ref{tab:pred} is presented at the Centre de Données at Strasbourg (CDS). Here only a 
part of the table is presented for illustration purposes. The table includes 
quantum numbers, calculated transition frequencies and corresponding uncertainties, 
the base 10 logarithm of the integrated transition intensity in the units of JPL and 
CDMS catalogs, nm$^2$MHz, and the lower state energy in cm$^{-1}$. Rotational 
$Q_{rot}$ and vibrational $Q_{v}$ partition functions used for the calculation 
of spectral predictions are given in Table~\ref{tab:qf}. The total partition 
function is thus $Q_{tot} = Q_{rot}\times Q_{v}$. The vibrational contribution 
to the partition function is calculated as ${\displaystyle Q_{v}= \prod_{i}\left(1-e^{-E_{i}/k_{B}T}\right)^{-1}}$, 
see equation 3.60 from \cite{gordy1984}, where $E_{i}$ is the energy 
of $i$-th vibrational mode. The $E_{i}$ values for the $\nu_{12}$ and $\nu_{9}$ 
modes were taken from the relative intensity measurements by \citet{sugisaki1974}, 
and for all other vibrational modes from the theoretical calculations using 
the vibrational self-consistent field method and taking second-order perturbative 
energy correction into account \citep{kowal2006}.

The nuclear quadrupole hyperfine structure was not taken into account 
in spectral predictions. In the previous study \citep{sugisaki1974}, 
a limited number of resolved hyperfine transitions did not allow 
an accurate determination of nuclear quadrupole coupling parameters 
because two different but relatively close sets of such parameters were obtained. 
Therefore the nuclear quadrupole hyperfine structure in the rotational 
spectrum of thioformamide may be a topic of future studies. We also verified 
that the hyperfine structure may be ignored for the rotational lines used for 
the search of thioformamide in the ISM when the spectral resolution of the 
ReMoCA survey is taken into account. For this purpose, we calculated two sets of spectral 
predictions of the ground-state rotational transitions of thioformamide using the 
two sets of nuclear hyperfine parameters from \citet{sugisaki1974}. 
In the two sets of predictions, the strongest $a$-type 
rotational transitions exhibit relatively small hyperfine splittings below 
the spectral resolution of the ReMoCA survey.

\section{Search for thioformamide toward Sgr~B2(N)}
\label{s:astro}

\subsection{Observations}
\label{ss:observations}

We used data acquired with the imaging spectral line survey ReMoCA performed with ALMA toward 
Sgr~B2(N). The observational setup and data 
reduction of the ReMoCA survey were described in \citet{Belloche19}. In short, 
the frequency range from 84.1~GHz to 114.4~GHz was fully covered with a 
spectral resolution of 488~kHz (1.7 to 1.3~km~s$^{-1}$). The interferometric
observations achieved an angular resolution (half-power beam width, HPBW) varying between 
$\sim$0.3$\arcsec$ and $\sim$0.8$\arcsec$, with a median value of 
0.6$\arcsec$. This corresponds to
$\sim$4900~au at the distance of Sgr~B2 \citep[8.2~kpc,][]{Reid19}. The rms 
sensitivity ranged between 0.35~mJy~beam$^{-1}$ and 1.1~mJy~beam$^{-1}$, with a 
median value of 0.8~mJy~beam$^{-1}$. The phase center was set to 
($\alpha, \delta$)$_{\rm J2000}$= 
($17^{\rm h}47^{\rm m}19{\fs}87, -28^\circ22'16{\farcs}0$), a position that is 
halfway between the two main hot molecular cores Sgr~B2(N1) and Sgr~B2(N2)
that are separated by 4.9$\arcsec$ or $\sim$0.2~pc in projection onto the 
plane of the sky. 

Here we analyze the spectra toward Sgr~B2(N2) at 
($\alpha, \delta$)$_{\rm J2000}$= 
($17^{\rm h}47^{\rm m}19{\fs}860$, $-28^\circ22\arcmin13{\farcs}40$)
and toward the offset position Sgr~B2(N1S) defined by \citet{Belloche19}. This 
offset position is located at ($\alpha, \delta$)$_{\rm J2000}$= 
($17^{\rm h}47^{\rm m}19{\fs}870$, $-28^\circ22\arcmin19{\farcs}48$), about 
1$\arcsec$ to the south of Sgr~B2(N1). It was chosen for two reasons: the
velocity dispersion (full width at half maximum, FWHM) of its molecular emission lines is small 
($\sim$5~km~s$^{-1}$), thus reducing the spectral confusion, and its continuum 
emission has a lower optical depth than the partially optically thick 
emission toward the peak position of the main hot core Sgr~B2(N1), thus 
allowing us to peer deeper into the molecular line emission region. 
Compared to \citet{Belloche19}, we used a new version of our data set for which
we have improved the separation of the continuum and line emission 
\citep[see][]{Melosso20}.

We modeled the spectra of Sgr~B2(N1S) and Sgr~B2(N2) with the software Weeds 
\citep[][]{Maret11} under the assumption of local thermodynamic equilibrium 
(LTE), which is appropriate given the high densities that characterize the 
regions where hot-core emission is detected in Sgr~B2(N) 
\citep[$>1 \times 10^{7}$~cm$^{-3}$, see][]{Bonfand19}. We derived a best-fit 
synthetic spectrum of each molecule separately, and then added the 
contributions of all identified molecules together. Each species was modeled 
with a set of five parameters: size of the emitting region ($\theta_{\rm s}$), 
column density ($N$), temperature ($T_{\rm rot}$), line width ($\Delta V$), and 
velocity offset ($V_{\rm off}$) with respect to the assumed systemic velocity 
of the source, $V_{\rm sys}=62$~km~s$^{-1}$ for Sgr~B2(N1S) and 
$V_{\rm sys}= 74$~km~s$^{-1}$ for Sgr~B2(N2).

\begin{table*}[!ht]
 \begin{center}
 \caption{
 Parameters of our best-fit LTE model of selected complex organic molecules toward Sgr~B2(N1S), and upper limit for thioformamide.
}
 \label{t:coldens_n1s}
 \vspace*{-1.2ex}
 \begin{tabular}{lcrccccccr}
 \hline\hline
 \multicolumn{1}{c}{Molecule} & \multicolumn{1}{c}{Status\tablefootmark{a}} & \multicolumn{1}{c}{$N_{\rm det}$\tablefootmark{b}} & \multicolumn{1}{c}{$\theta_{\rm s}$\tablefootmark{c}} & \multicolumn{1}{c}{$T_{\mathrm{rot}}$\tablefootmark{d}} & \multicolumn{1}{c}{$N$\tablefootmark{e}} & \multicolumn{1}{c}{$F_{\rm vib}$\tablefootmark{f}} & \multicolumn{1}{c}{$\Delta V$\tablefootmark{g}} & \multicolumn{1}{c}{$V_{\mathrm{off}}$\tablefootmark{h}} & \multicolumn{1}{c}{$\frac{N_{\rm ref}}{N}$\tablefootmark{i}} \\ 
  & & & \multicolumn{1}{c}{\small ($''$)} & \multicolumn{1}{c}{\small (K)} & \multicolumn{1}{c}{\small (cm$^{-2}$)} & & \multicolumn{1}{c}{\small (km~s$^{-1}$)} & \multicolumn{1}{c}{\small (km~s$^{-1}$)} & \\ 
 \hline
 CH$_3$OH, $\varv=0$$^\star$ & d & 35 &  2.0 &  230 &  2.0 (19) & 1.00 & 5.0 & $0.2$ &       1 \\ 
 \hspace*{8ex} $\varv_{\rm t}=1$ & d & 17 &  2.0 &  230 &  2.0 (19) & 1.00 & 5.0 & $0.2$ &       1 \\ 
 \hspace*{8ex} $\varv_{\rm t}=2$ & d & 4 &  2.0 &  230 &  2.0 (19) & 1.00 & 5.0 & $0.2$ &       1 \\ 
 \hspace*{8ex} $\varv_{\rm t}=3$ & t & 1 &  2.0 &  230 &  2.0 (19) & 1.00 & 5.0 & $0.2$ &       1 \\ 
 $^{13}$CH$_3$OH, $\varv=0$ & d & 10 &  2.0 &  230 &  1.2 (18) & 1.00 & 5.0 & $0.2$ &      17 \\ 
 \hspace*{9.5ex} $\varv_{\rm t}=1$ & d & 6 &  2.0 &  230 &  1.2 (18) & 1.00 & 5.0 & $0.2$ &      17 \\ 
 CH$_3$$^{18}$OH, $\varv=0$ & t & 1 &  2.0 &  230 &  2.0 (17) & 1.00 & 5.0 & $0.2$ &     100 \\ 
 \hspace*{9.5ex} $\varv_{\rm t}=1$ & n & 0 &  2.0 &  230 &  2.0 (17) & 1.00 & 5.0 & $0.2$ &     100 \\ 
\hline 
 CH$_3$SH, $\varv=0$ & d & 6 &  2.0 &  250 &  5.5 (17) & 1.00 & 5.0 & $0.0$ &      36 \\ 
 \hspace*{7.3ex} $\varv_{\rm t} = 1$ & t & 0 &  2.0 &  250 &  5.5 (17) & 1.00 & 5.0 & $0.0$ &      36 \\ 
 \hspace*{7.3ex} $\varv_{\rm t} = 2$ & n & 0 &  2.0 &  250 &  5.5 (17) & 1.00 & 5.0 & $0.0$ &      36 \\ 
\hline 
 NH$_2$CHO\tablefootmark{(j)}$^\star$ & d & 34 &  2.0 &  160 &  2.9 (18) & 1.09 & 6.0 & $0.0$ &       1 \\ 
 NH$_2$CHS, $\varv=0$ & n & 0 &  2.0 &  160 & $<$  4.2 (15) & 1.05 & 6.0 & $0.0$ & $>$     701 \\ 
\hline 
 \end{tabular}
 \end{center}
 \vspace*{-2.5ex}
 \tablefoot{
 \tablefoottext{a}{d: detection, t: tentative detection, n: nondetection.}
 \tablefoottext{b}{Number of detected lines \citep[conservative estimate, see Sect.~3 of][]{Belloche16}. One line of a given species may mean a group of transitions of that species that are blended together.}
 \tablefoottext{c}{Source diameter (FWHM).}
 \tablefoottext{d}{Rotational temperature.}
 \tablefoottext{e}{Total column density of the molecule. $x$ ($y$) means $x \times 10^y$. An identical value for all listed vibrational/torsional states of a molecule means that LTE is an adequate description of the vibrational/torsional excitation.}
 \tablefoottext{f}{Correction factor that was applied to the column density to account for the contribution of vibrationally excited states, in the cases where this contribution was not included in the partition function of the spectroscopic predictions.}
 \tablefoottext{g}{Linewidth (FWHM).}
 \tablefoottext{h}{Velocity offset with respect to the assumed systemic velocity of Sgr~B2(N1S), $V_{\mathrm{sys}} = 62$ km~s$^{-1}$.}
 \tablefoottext{i}{Column density ratio, with $N_{\rm ref}$ the column density of the previous reference species marked with a $\star$.}
 \tablefoottext{j}{The parameters were derived from the ReMoCA survey by \citet{Belloche19}.}
 }
 \end{table*}

\begin{table*}[!ht]
 \begin{center}
 \caption{
 Parameters of our best-fit LTE model of selected complex organic molecules toward Sgr~B2(N2) and upper limit for thioformamide.
}
 \label{t:coldens_n2}
 \vspace*{-1.2ex}
 \begin{tabular}{lcrccccccr}
 \hline\hline
 \multicolumn{1}{c}{Molecule} & \multicolumn{1}{c}{Status\tablefootmark{(a)}} & \multicolumn{1}{c}{$N_{\rm det}$\tablefootmark{(b)}} & \multicolumn{1}{c}{$\theta_{\rm s}$\tablefootmark{(c)}} & \multicolumn{1}{c}{$T_{\mathrm{rot}}$\tablefootmark{(d)}} & \multicolumn{1}{c}{$N$\tablefootmark{(e)}} & \multicolumn{1}{c}{$F_{\rm vib}$\tablefootmark{(f)}} & \multicolumn{1}{c}{$\Delta V$\tablefootmark{(g)}} & \multicolumn{1}{c}{$V_{\mathrm{off}}$\tablefootmark{(h)}} & \multicolumn{1}{c}{$\frac{N_{\rm ref}}{N}$\tablefootmark{i}} \\ 
  & & & \multicolumn{1}{c}{\small ($''$)} & \multicolumn{1}{c}{\small (K)} & \multicolumn{1}{c}{\small (cm$^{-2}$)} & & \multicolumn{1}{c}{\small (km~s$^{-1}$)} & \multicolumn{1}{c}{\small (km~s$^{-1}$)} & \\ 
 \hline
 CH$_3$OH\tablefootmark{(j)}\tablefootmark{(k)}$^\star$ & d & 60 &  1.4 &  160 &  4.0 (19) & 1.00 & 5.4 & $-0.2$ &       1 \\ 
 CH$_3$SH\tablefootmark{(j)} & d & 13 &  1.4 &  180 &  3.4 (17) & 1.00 & 5.4 & $-0.5$ &     118 \\ 
\hline 
 NH$_2$CHO\tablefootmark{(j)}\tablefootmark{(k)}$^\star$ & d & 43 &  0.8 &  200 &  2.6 (18) & 1.17 & 5.5 & $0.2$ &       1 \\ 
 NH$_2$CHS, $\varv=0$ & n & 0 &  0.8 &  200 & $<$  2.8 (15) & 1.12 & 5.5 & $0.2$ & $>$     919 \\ 
\hline 
 \end{tabular}
 \end{center}
 \vspace*{-2.5ex}
 \tablefoot{
 \tablefoottext{a}{d: detection, n: nondetection.}
 \tablefoottext{b}{Number of detected lines \citep[conservative estimate, see Sect.~3 of][]{Belloche16}. One line of a given species may mean a group of transitions of that species that are blended together.}
 \tablefoottext{c}{Source diameter (FWHM).}
 \tablefoottext{d}{Rotational temperature.}
 \tablefoottext{e}{Total column density of the molecule. $x$ ($y$) means $x \times 10^y$.}
 \tablefoottext{f}{Correction factor that was applied to the column density to account for the contribution of vibrationally excited states, in the cases where this contribution was not included in the partition function of the spectroscopic predictions.}
 \tablefoottext{g}{Linewidth (FWHM).}
 \tablefoottext{h}{Velocity offset with respect to the assumed systemic velocity of Sgr~B2(N2), $V_{\mathrm{sys}} = 74$ km~s$^{-1}$.}
 \tablefoottext{i}{Column density ratio, with $N_{\rm ref}$ the column density of the previous reference species marked with a $\star$.}
 \tablefoottext{j}{The parameters were derived from the EMoCA survey by \citet{Mueller16} and \citet{Belloche17}.}
 \tablefoottext{k}{For CH$_3$OH and NH$_2$CHO, we report the parameters derived from the vibrationally excited states $\varv_{\rm t}=1$ and $\varv_{12}=1$, respectively.}
 }
 \end{table*}

\begin{table}
 \begin{center}
 \caption{ Rotational temperatures derived from population diagrams toward Sgr~B2(N1S).}
 \label{t:popfit_n1s}
 \vspace*{0.0ex}
 \begin{tabular}{lll}
 \hline\hline
 \multicolumn{1}{c}{Molecule} & \multicolumn{1}{c}{States\tablefootmark{a}} & \multicolumn{1}{c}{$T_{\rm fit}$\tablefootmark{b}} \\ 
  & & \multicolumn{1}{c}{\small (K)} \\ 
 \hline
CH$_3$OH & $\varv=0$, $\varv_{\rm t}=1$, $\varv_{\rm t}=2$, $\varv_{\rm t}=3$ & 234.9 (7.7) \\ 
$^{13}$CH$_3$OH & $\varv=0$, $\varv_{\rm t}=1$ &   232 (21) \\ 
CH$_3$$^{18}$OH & $\varv=0$, $\varv_{\rm t}=1$ &   195 (52) \\ 
\hline 
CH$_3$SH & $\varv=0$, $\varv_{\rm t}=1$ &   335 (90) \\ 
\hline 
 \end{tabular}
 \end{center}
 \vspace*{-2.5ex}
 \tablefoot{
 \tablefoottext{a}{Vibrational states that were taken into account to fit the population diagram.}
 \tablefoottext{b}{The standard deviation of the fit is given in parentheses. As explained in Sect.~3 of \citet{Belloche16} and Sect.~4.4 of \citet{Belloche19}, this uncertainty is purely statistical and should be viewed with caution. It may be underestimated.}
 }
 \end{table}

\subsection{Nondetection of thioformamide}
\label{ss:detection}

\begin{figure*}[!t]
\centerline{\resizebox{0.88\hsize}{!}{\includegraphics[angle=0]{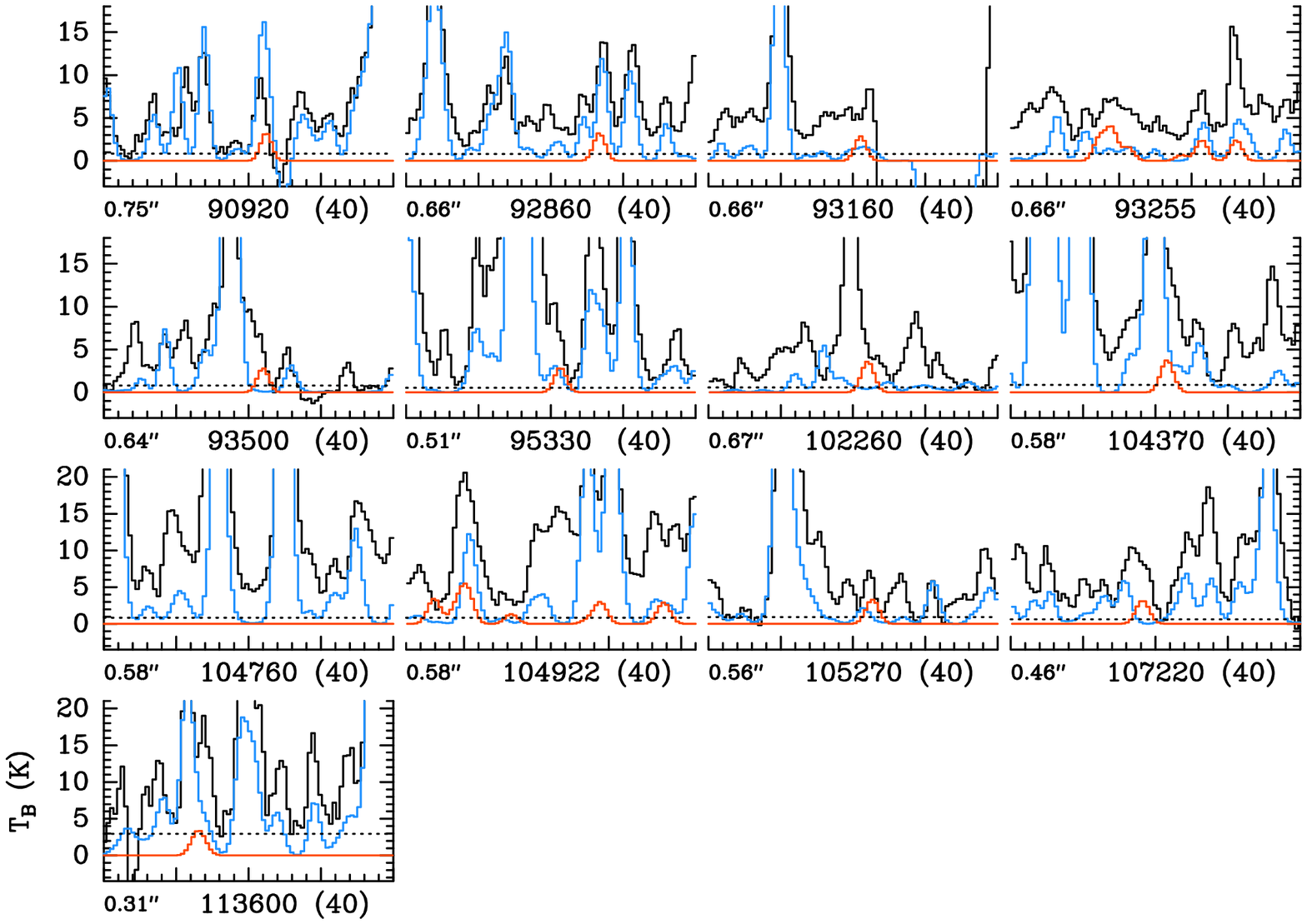}}}
\caption{Transitions of NH$_2$CHS, $\varv = 0$ covered by our ALMA 
survey. The synthetic spectrum of NH$_2$CHS, $\varv = 0$ we used to derive 
the upper limit to its column density is displayed in red and overlaid on the 
observed spectrum of Sgr~B2(N1S) shown in black. The blue synthetic spectrum 
contains the contributions from all molecules identified in our survey so far, 
but not from the species shown in red. 
The central frequency and width are indicated in MHz below each panel. The
angular resolution (HPBW) is also indicated. The $y$-axis is labeled in
brightness temperature units (K). The dotted line indicates the $3\sigma$ 
noise level.}
\label{f:spec_nh2chs_ve0_n1s}
\end{figure*}

\begin{figure*}[!t]
\centerline{\resizebox{0.88\hsize}{!}{\includegraphics[angle=0]{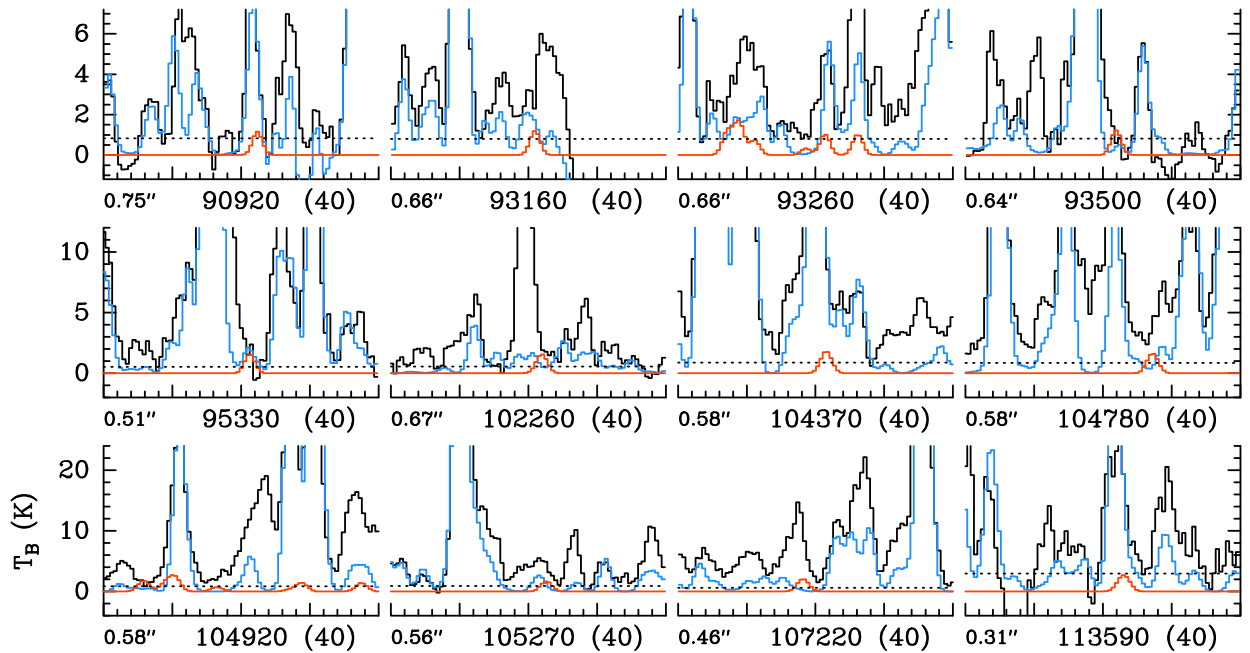}}}
\caption{Same as Fig.~\ref{f:spec_nh2chs_ve0_n1s}, but for Sgr~B2(N2).}
\label{f:spec_nh2chs_ve0_n2}
\end{figure*}

We searched for rotational emission of thioformamide toward Sgr~B2(N1S) and
Sgr~B2(N2) by comparing the ReMoCA spectra to LTE synthetic spectra of 
thioformamide obtained with the parameters derived for formamide, 
NH$_2$CHO, by \citet{Belloche19} and \citet{Belloche17}, respectively, keeping 
only the column density of thioformamide as a free parameter. We found no 
evidence for emission of thioformamide toward either source.
Figures~\ref{f:spec_nh2chs_ve0_n1s} and \ref{f:spec_nh2chs_ve0_n2} illustrate
the nondetection toward Sgr~B2(N1S) and Sgr~B2(N2), respectively, and the 
upper limits to the column density of thioformamide are indicated in 
Tables~\ref{t:coldens_n1s} and \ref{t:coldens_n2}, respectively, along with
the parameters previously obtained for formamide.

\subsection{Comparison to other molecules}
\label{ss:otherCOMs}

In order to place the nondetection of thioformamide into an astrochemical 
context, we wished to compare the upper limits on its column density to the 
column densities of other complex organic molecules. In addition to formamide, 
we chose methanol, CH$_3$OH, and methanethiol, CH$_3$SH, which were both
detected toward both hot cores \citep[][]{Belloche13,Mueller16}.
Table~\ref{t:coldens_n2} lists the parameters obtained for methanol and 
methanethiol by \citet{Mueller16} toward Sgr~B2(N2) on the basis of the EMoCA 
survey, which was also performed with ALMA \citep[][]{Belloche16}.

We derived the column densities of methanol and methanethiol toward 
Sgr~B2(N1S) using the ReMoCA survey. In order to obtain the LTE parameters of
methanol, we also modeled the emission of its isotopologs, $^{13}$CH$_3$OH
and CH$_3^{18}$OH. We used the spectroscopic entries available in the CDMS
database to compute the synthetic spectra: version 3 of entry 32504 for 
CH$_3$OH, version 2 of entry 33502 for $^{13}$CH$_3$OH, version 1 of entry 
34504 for CH$_3^{18}$OH, and version 2 of entry 48510 for CH$_3$SH.

Our best-fit synthetic spectra are shown in 
Figs.~\ref{f:spec_ch3oh_ve0_n1s}-\ref{f:spec_ch3oh_18o_ve1_n1s} for methanol
and its isotopologs, and in Figs.~\ref{f:spec_ch3sh_ve0_n1s} and 
\ref{f:spec_ch3sh_ve1_n1s} for methanethiol. The parameters of these
LTE models are listed in Table~\ref{t:coldens_n1s}. Methanol is detected in 
its vibrational ground state and in its first two vibrational
states (Figs.~\ref{f:spec_ch3oh_ve0_n1s}--\ref{f:spec_ch3oh_ve2_n1s}). It is 
only tentatively detected in the third vibrational state 
(Fig.~\ref{f:spec_ch3oh_ve3_n1s}). $^{13}$CH$_3$OH is detected in both the 
ground and first vibrational states (Figs~\ref{f:spec_ch3oh_13c_ve0_n1s} and
\ref{f:spec_ch3oh_13c_ve1_n1s}). There are hints of emission from 
CH$_3^{18}$OH (Figs.~\ref{f:spec_ch3oh_18o_ve0_n1s} and 
\ref{f:spec_ch3oh_18o_ve1_n1s}), but because only one transition is not much 
contaminated by emission from other species, the identification of this 
isotopolog and the derived column density are uncertain. The true column 
density might well be lower. For certain \ce{CH3OH} lines, the 
actually observed ReMoCA spectrum shows much weaker emission than predicted by the 
LTE model or even absorption. This is due to absorption from the extended envelope 
of Sgr~B2(N) and/or due to peculiarities in the excitation of methanol that lead to 
non-LTE effects. These may result in inversion for some lines and in anti-inversion
(``overcooling'') for others; for a discussion, see Section 5.4 of \citet{Belloche13}. 

Methanethiol is clearly detected in its vibrational ground state 
(Fig.~\ref{f:spec_ch3sh_ve0_n1s}). Three transitions from within 
$\varv_{\rm t} = 1$ match the observed lines well 
(Fig.~\ref{f:spec_ch3sh_ve1_n1s}), but they account for only $\sim$50\% of the 
detected signal, the rest being emitted by other species included in our full 
model. This is not sufficient to claim a firm detection of this state, but we 
consider it as tentatively detected. Emission from several transitions of 
$\varv_{\rm t}=2$ significantly contributes to the detected signal 
(Fig.~\ref{f:spec_ch3sh_ve2_n1s}), but no line can be firmly identified, and 
so no detection of this state can be claimed. However, we include it in our 
full model of  the Sgr~B(N1S) spectrum.


We used the lines that are not too strongly contaminated to produce 
population diagrams for methanol, its isotopologs, and methanethiol.
These diagrams are 
shown in Figs.~\ref{f:popdiag_ch3oh_n1s}--\ref{f:popdiag_ch3sh_n1s}. The 
rotational temperatures formally derived from fits to these population 
diagrams are reported in Table~\ref{t:popfit_n1s}. We adopted a temperature of 
230~K to model the emission of methanol and its isotopologs. The temperature 
of methanethiol is not well constrained, and we set it to 250~K, which 
gives a reasonable fit to the emission of its ground and first vibrational 
states.

\begin{figure}[!t]
\centerline{\resizebox{1.0\hsize}{!}{\includegraphics[angle=0]{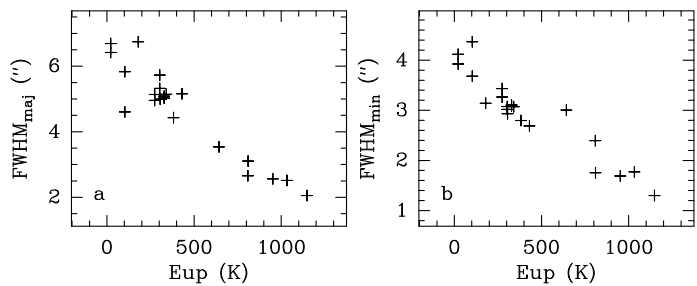}}}
\caption{Deconvolved emission size of uncontaminated methanol transitions as a 
function of upper-level energy. Panels \textbf{a} and \textbf{b} show the 
major and minor axes (FWHM), respectively.}
\label{f:size_ch3oh_n1s}
\end{figure}

The LTE synthetic spectra of Sgr~B2(N1S) were computed assuming a diameter 
(FWHM) of 2$\arcsec$ for the emission of methanol and methanethiol, as was 
done for formamide and other species reported by \citet{Belloche19} toward 
Sgr~B2(N1S). This size is approximately three times larger than the angular 
resolution of the ReMoCA survey, and thus its exact value does not affect the derivation of the column density significantly. A two-dimensional Gaussian fit to 
the integrated intensity maps of the least contaminated transitions of 
methanethiol yields a mean size of $\sim$2.2$\arcsec$ with an rms 
dispersion of $\sim$0.3$\arcsec$, consistent with our size assumption. The 
emission of higher-energy transitions of CH$_3$OH and $^{13}$CH$_3$OH also 
has a typical size of about 2$\arcsec$, while lower-energy transitions show
more extended emission up to $\sim$5$\arcsec$ (Fig.~\ref{f:size_ch3oh_n1s}). 
This indicates that in addition to the compact hot-core component, there is also 
more extended emission of methanol in the envelope of the main hot core. This 
probably explains why the most optically
thick transitions are not well reproduced by our synthetic model: their peak 
intensities are overestimated (see Fig.~\ref{f:spec_ch3oh_ve0_n1s}). Our model 
does not account for extended methanol emission at lower densities in the 
envelope, which likely acts as an opaque screen at the frequencies of optically 
thick transitions, thus attenuating their peak intensities.

\section{Discussion}
\label{s:discussion}

No emission of thioformamide was detected in the ReMoCA spectra of Sgr~B2(N1S) 
and Sgr~B2(N2). The column density upper limits derived from our LTE 
analysis indicate that thioformamide is at least $\sim$700 and $\sim$900 
times less abundant than formamide in Sgr~B2(N1S) and Sgr~B2(N2), 
respectively. This is at first sight surprising when we compare it to methanethiol, 
which is only $\sim$40 and 120 times less abundant than methanol in
these sources, respectively. We examine below which chemical processes might 
be responsible for this large difference.


Astrochemical kinetic modeling of sulfur-bearing species has mostly been 
limited to molecules only as large as H$_2$CS \citep[e.g.,][]{Charnley97}, but 
\citet{Mueller16} constructed a chemical network for methanethiol and 
ethanethiol that was based mostly on grain-surface chemistry (while also 
including the necessary gas-phase destruction mechanisms). To our knowledge, 
no networks exist for the production of interstellar thioformamide. It is 
therefore not possible to compare the observational results directly with a 
chemical model. However, guided by the modeling results for related molecules, 
we may draw qualified conclusions as to the origin of the different behavior 
of formamide/thioformamide versus methanol/methanethiol.

Recently, \citet{Jin20} presented rate equation-based chemical kinetic models 
that included a range of new, nondiffusive reaction mechanisms on cold 
dust-grain surfaces. In the usual scenario, the chemistry at low temperatures 
is dominated by mobile H atoms, which diffuse on the grain or ice surface until 
they meet and react with some relatively immobile species; this might be, for 
example, a heavier atom adsorbed from the gas such as O, a molecule such as CO, 
or a radical produced by a previous surface reaction. The inclusion of 
nondiffusive processes in the models occasionally allows surface radicals to 
react with each other without significant diffusion being required (which is 
largely prohibited at low temperatures). When produced on grain surfaces 
through any chemical reaction, radicals may sometimes be formed in close 
proximity to each other, either in direct contact or within a short distance, 
such that they may occasionally react as soon as either one is formed in the 
presence of the other. The exothermicity of the initiating reaction would also 
allow some degree of surface diffusion to take place, allowing nearby but 
noncontiguous species to meet after the initiating reaction. \citet{Jin20} 
referred to this overall nondiffusive process as the three-body (3-B) 
reaction mechanism. (Other nondiffusive mechanisms also exist that are 
initiated by other processes). This 3-B mechanism has been explored 
experimentally over the past several years and appears to be effective in 
producing complex organic molecules, especially those derived from the 
hydrogenation of CO \citep[e.g.,][]{Fedoseev17}. This general mechanism is 
usually present automatically in microscopic kinetic Monte Carlo models of 
grain-surface chemistry, which consider the explicit positions of surface 
species, but must be explicitly coded into models that are based on rate 
equations.

\citet{Garrod20} applied the nondiffusive processes investigated by 
\citet{Jin20} to hot-core chemical models; as in previous work, the physical 
evolution involves a cold-collapse stage during which the dust temperature 
falls from (in this model) approximately 15~K down to 8~K as the visual 
extinction rises. The collapse stage is followed by a time-dependent warm-up 
of the gas and dust, from 8 to 400~K. In their ``final'' model, encompassing 
all of the new mechanisms and updates, \citet{Garrod20} find that formamide is 
produced almost entirely during the cold stage, as the result of both 
diffusive and nondiffusive reaction processes on the dust or ice surfaces, with 
those molecules then becoming embedded in the ice mantles and returning to the 
gas phase when high temperatures ($>$100~K) are reached. The critical 
nondiffusive step in this formation process is the reaction between HCO and
 NH, to produce HNCHO, which may be further hydrogenated by diffusive 
H-addition to produce NH$_2$CHO. The radicals HCO and NH may be formed by the 
diffusive addition of H to CO and N, respectively, which may on occasion react 
spontaneously when formed nearby (i.e., the 3-B reaction).

\begin{figure}[!t]
\centerline{\resizebox{1.0\hsize}{!}{\includegraphics[angle=0]{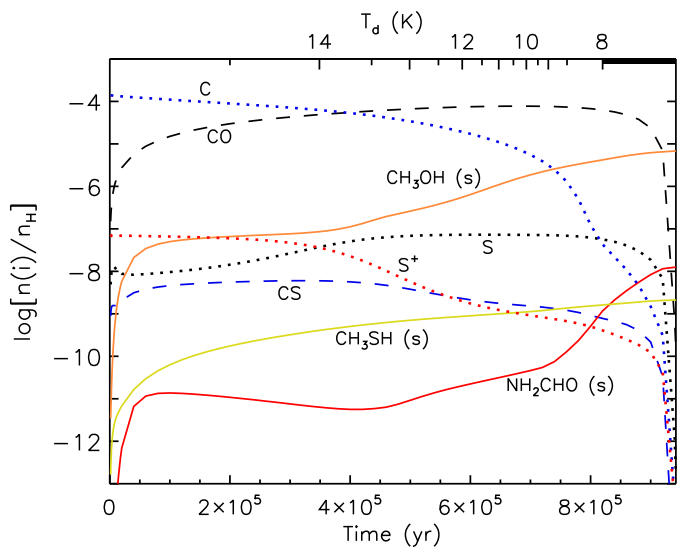}}}
\caption{Fractional abundances with respect to total hydrogen for selected 
gas- and solid-phase species during the cold-collapse stage of hot core 
evolution based on data from the ``final'' model of \citet{Garrod20}.
Solid-phase species are marked ``(s)'' and are plotted with a solid line. The 
dust temperature is indicated on the top axis and has a lower limit of 8~K.}
\label{f:chemmodel}
\end{figure}

Figure~\ref{f:chemmodel} shows results from the \citet{Garrod20} model for a 
selection of species related to thioformamide and methanethiol for the cold 
collapse phase (during which most solid-phase formamide is produced). The 
collapse from a density of $3 \times 10^3$ to $2 \times 10^8$~cm$^{-3}$ takes 
approximately $9.4 \times 10^5$~yr. Methanethiol is explicitly included in the 
chemical model, using the sulfur network presented by \citet{Mueller16}, with 
the same sulfur elemental abundance of $8 \times 10^{-8}n_{\mathrm{H}}$.

The fractional abundance of solid-phase (marked ``(s)'') 
methanethiol gradually grows during the collapse, reaching a final abundance 
of a few $10^{-9}$. An abundance of this order is reached at around 
$7 \times 10^5$~yr. As with methanol forming from grain surface CO, 
methanethiol is formed by repeated H addition to CS. The CS is accreted from 
the gas phase, forming mainly as the result of reactions of S$^+$ with CH 
(whose abundance tracks with that of gas-phase atomic C), leading to CS$^+$. 
Reaction with H$_2$ to give HCS$^+$, followed by recombination with electrons completes 
the conversion into CS.

Solid-phase formamide abundance is also shown in Fig.~\ref{f:chemmodel}; it 
should be noted that the above-described 3-B process for formamide only leads 
to substantial formation of this molecule at late times in the model, when 
temperatures are very low ($< \sim$9~K) and densities high. The slower surface 
diffusion of atomic H and its reduced adsorption from the gas phase under 
these conditions allow the radicals to build up a somewhat greater surface 
coverage, which increases the influence of the 3-B process.

In a similar way to formamide, thioformamide might also be expected to form at
low temperatures or late times through the 3-B process through the reaction 
HCS + NH $\rightarrow$ HNCHS, followed by diffusive hydrogen addition to give 
NH$_2$CHS. However, in this case, the necessary gas-phase precursor species, 
CS, is much lower in abundance at late times when formamide or thioformamide 
production should be highest. Gas-phase CS abundance begins to fall relative 
to CO, starting at around $4 \times 10^5$ yr, as the result of the switch-over 
from S$^+$ to neutral S as the dominant sulfur carrier in the gas phase. The 
drop in S$^+$ together with the gradual fall in atomic C and CH abundances
ultimately reduces the CS abundance, meaning that toward the end time in the 
model, production of NH$_2$CHS would be much weaker (relative to NH$_2$CHO) 
than if the earlier gas-phase CS abundance had been preserved. The gas-phase 
ratio of CS/CO at a time $4 \times 10^5$ yr is approximately 
$1.1 \times 10^{-4}$, while at a time of $9 \times 10^5$ yr, when NH$_2$CHO 
production is at its peak, this ratio is about $9.2 \times 10^{-6}$, which is a 
decrease of about 12 times. At yet later times, the ratio continues to fall.

We therefore suggest that this fall in gas-phase CS at just the period when 
surface production of NH$_2$CHS should be most efficient is the underlying 
cause for the weaker production of NH$_2$CHS as compared with the production of CH$_3$SH. The latter can form 
more consistently at earlier times and over a broader range of temperatures because it is produced through direct H-addition and is independent of the 
more restrictive 3-B process.

As a point of comparison, we note that the solid-phase CH$_3$SH 
abundance presented in Fig.~\ref{f:chemmodel} is similar to that found by 
\citet{Mueller16} because it is formed by the same standard process of 
diffusive H-addition. We\ also note that the reduced elemental 
sulfur abundance that is used in both models, while typical for such studies, could 
produce a different result from a model using the much higher solar or 
diffuse-cloud S abundances. However, the ratio of methanethiol to methanol in the model is about two orders of magnitude lower than the observational value, therefore a higher elemental sulfur abundance in the models might be justified on this basis. The dominant form of this sulfur in dense 
interstellar regions is still uncertain and might affect the chemistry 
described above.

It is possible that some of the species discussed here may be formed during 
the warm-up stage, likely through radical-radical reactions driven by 
cosmic-ray-induced UV photodissociation. Based on the \citet{Garrod20} models, we do 
not expect this to make a substantial contribution to the relative abundances 
of methanol and formamide and the corresponding 
sulfur derivatives. It has also 
been suggested that gas-phase mechanisms might play a role in the production 
of formamide in hot cores \citep[e.g.,][]{Barone15} through the reaction 
NH$_2$ + H$_2$CO $\rightarrow$ NH$_2$CHO + H. However, again, the models of 
\citet{Garrod20} suggest that the effect of such processes, even using the 
rates provided by \citet{Skouteris17}, is slight compared to grain-surface 
production. It is unknown whether a similar process might contribute 
substantially to the production of thioformamide in the gas phase.

A more robust investigation of the possible production of interstellar 
thioformamide would of course involve the use of a dedicated chemical network 
for that and related species; we hope to carry out such a study in the future.

\section{Conclusions}
\label{s:conclusions}

We have presented a comprehensive study of the rotational spectrum of thioformamide 
that includes characterization of the parent as well the other most abundant 
isotopic species. On the basis of the results of this study, accurate frequency 
predictions of the ground-state rotational spectra of thioformamide were 
obtained for the transitions involving levels with $J \leq 90$ and $K_{a} \leq 35$ 
and in the frequency range at least up to 1~THz. These predictions have enabled 
the first search for thioformamide in the interstellar medium. 
The main conclusions of this search are listed below.
\begin{enumerate}

\item We report the nondetection of thioformamide toward the hot cores 
Sgr~B2(N1S) and Sgr~B2(N2) with ALMA. The sensitive upper limits indicate that 
thioformamide is at least nearly three orders of magnitude less abundant than 
formamide, which is surprising because only two orders of magnitude 
 separate methanethiol and methanol in these sources.

\item Models indicate that the production of formamide, and thus 
thioformamide, may rely on nondiffusive surface chemistry between radicals, 
unlike methanethiol, which may be formed on grain surfaces by the repetitive
hydrogenation of CS accreted from the gas phase. Optimal conditions for 
nondiffusive production of formamide or thioformamide occur at late times or low 
temperatures during the prestellar phase, when the abundance of gas-phase CS is depressed with respect to 
CO. This decreases the availability of the HCS radical on grains versus HCO, and 
thus reduces the ratio of the S:O content. The model results therefore provide a plausible explanation for the surprising observational result reported above.
\end{enumerate}

\begin{acknowledgements}
The present investigations were supported by the CNES and by the French Programme 
National ``Physique et Chimie du Milieu Interstellaire'' (PCMI). J.~C.~G. thanks 
the Centre National d’Etudes Spatiales (CNES) for a grant.
This paper makes use of the following ALMA data: 
ADS/JAO.ALMA\#2016.1.00074.S. 
ALMA is a partnership of ESO (representing its member states), NSF (USA), and 
NINS (Japan), together with NRC (Canada), NSC and ASIAA (Taiwan), and KASI 
(Republic of Korea), in cooperation with the Republic of Chile. The Joint ALMA 
Observatory is operated by ESO, AUI/NRAO, and NAOJ. The interferometric data 
are available in the ALMA archive at https://almascience.eso.org/aq/.
Part of this work has been carried out within the Collaborative
Research Centre 956, sub-project B3, funded by the Deutsche
Forschungsgemeinschaft (DFG) -- project ID 184018867.
RTG acknowledges support from the National Science Foundation through 
grant number AST 19-06489.

\end{acknowledgements}

\bibliographystyle{aa}
\bibliography{thiofmm}

\begin{appendix}
\label{appendix}

\section{Observed and predicted transitions of the ground-state rotational spectrum of thioformamide}

The spectral predictions given in Table~\ref{tab:pred} were calculated at $T=300$~K using a set of parameters from Table~\ref{tab:rot}, and a corresponding partition function value from Table~\ref{tab:qf}.

\begin{table*}
\centering
 \caption{Small part of the table that is available at the CDS, with assigned rotational transitions of the ground state of thioformamide (parent isotopolog).}  
 \label{tab:rottrans}   
\begin{tabular}{cccccccccc}
\hline\hline
 $J''$ & $K_a''$ & $K_c''$ & $J'$ & $K_a'$ & $K_c'$ & Measured frequency & Residual (MHz) &  Uncertainty & Weighted relative \\
  & & & & & & (MHz) & A-reduction &  (MHz) & intensity \\
\hline 
 35 & 23 & 12 & 34 & 23 & 11  & 409154.700 &   0.0130 &  0.030 & 0.5  \\
 35 & 23 & 13 & 34 & 23 & 12  & 409154.700 &   0.0130 &  0.030 & 0.5  \\
 35 & 24 & 11 & 34 & 24 & 10  & 409292.876 &   0.0012 &  0.050 & 0.5  \\
 35 & 24 & 12 & 34 & 24 & 11  & 409292.876 &   0.0012 &  0.050 & 0.5  \\
 35 &  4 & 31 & 34 &  4 & 30  & 409846.750 &   0.0025 &  0.030 &       \\
 35 &  3 & 32 & 34 &  3 & 31  & 413786.557 &  -0.0109 &  0.030 &       \\
 36 &  2 & 35 & 35 &  2 & 34  & 414200.369 &   0.0086 &  0.030 &       \\
 35 &  2 & 33 & 34 &  2 & 32  & 415506.771 &  -0.0096 &  0.100 &       \\
 37 &  1 & 37 & 36 &  1 & 36  & 415758.710 &  -0.0028 &  0.100 &       \\
 37 &  0 & 37 & 36 &  0 & 36  & 416065.293 &  -0.0067 &  0.100 &       \\
 \hline 
\end{tabular} 
 \end{table*} 

\begin{table*}
\centering
 \caption{Small part of the table that is available at the CDS, with predicted transitions of thioformamide in the ground vibrational state (parent isotopolog).}  
 \label{tab:pred}       
\begin{tabular}{cccccccccc}
\hline\hline                                                                                                                 
$J''$ & $K_a''$ & $K_c''$  & $J'$ & $K_a'$ & $K_c '$ & Calc. freq.  & Uncertainty & $\log I$       &  $E_l$     \\ 
     &        &         &       &         &          &   (MHz)      &    (MHz)    &   (nm$^2$MHz) &  cm$^{-1}$ \\
\hline                                                                                                                                           
  20 &  1 & 19 & 20 & 1 & 20  &  113907.1811  &  0.0693 &  -5.3017  &   81.3231 \\      
  31 &  2 & 29 & 31 & 2 & 30  &  114075.5111  &  0.0731 &  -5.2549  &  199.2715 \\      
  51 &  5 & 46 & 52 & 3 & 49  &  114914.8081  &  0.0532 &  -7.0851  &  559.4805 \\      
  36 &  2 & 34 & 37 & 0 & 37  &  115127.6928  &  0.1645 &  -7.0067  &  266.5758 \\      
  60 &  8 & 53 & 61 & 7 & 54  &  115583.6123  &  0.5572 &  -7.7911  &  827.0399 \\      
  10 &  0 & 10 &  9 & 0 &  9  &  115841.4912  &  0.0016 &  -3.1709  &   17.4442 \\      
  60 &  8 & 52 & 61 & 7 & 55  &  115855.7744  &  0.5570 &  -7.7890  &  827.0312 \\      
  18 &  1 & 17 & 18 & 0 & 18  &  116139.3026  &  0.0496 &  -6.1802  &   65.8358 \\      
  10 &  2 &  9 &  9 & 2 &  8  &  116406.8865  &  0.0014 &  -3.1998  &   24.9448 \\      
  10 &  5 &  6 &  9 & 5 &  5  &  116565.5620  &  0.0013 &  -3.3874  &   64.1034 \\          
 \hline                                                                       
\end{tabular}
\end{table*}  

\begin{table*}
\centering
 \caption{Rotational $Q_{rot}$ and vibrational $Q_{v}$ partition functions of thioformamide at various temperatures.}  
 \label{tab:qf} 
\begin{tabular}{ccc}
\hline\hline
Temperature (K) & $Q_{rot}$  & $Q_{v}$ \\
\hline
300       &  19204.4   &  1.4287 \\
225       &   12468.1   & 1.1741  \\
150       &   6784.2   &  1.0386 \\
75        &   2398.8   & 1.0007  \\
37.5  &   848.8    & 1.0000  \\
18.75 &   300.8    & 1.0000  \\
9.375 &   107.0     &1.0000   \\ 
\hline
\end{tabular}
\end{table*}

\section{Complementary figures: Spectra}
\label{a:spectra}

Figures~\ref{f:spec_ch3oh_ve0_n1s}--\ref{f:spec_ch3sh_ve2_n1s} show the rotational 
transitions of methanol, its isotopologs, and methanethiol that are 
covered by the \hbox{ReMoCA} survey and significantly contribute to the signal 
detected toward Sgr~B2(N1S). 

\begin{figure*}[!ht]
\centerline{\resizebox{0.88\hsize}{!}{\includegraphics[angle=0]{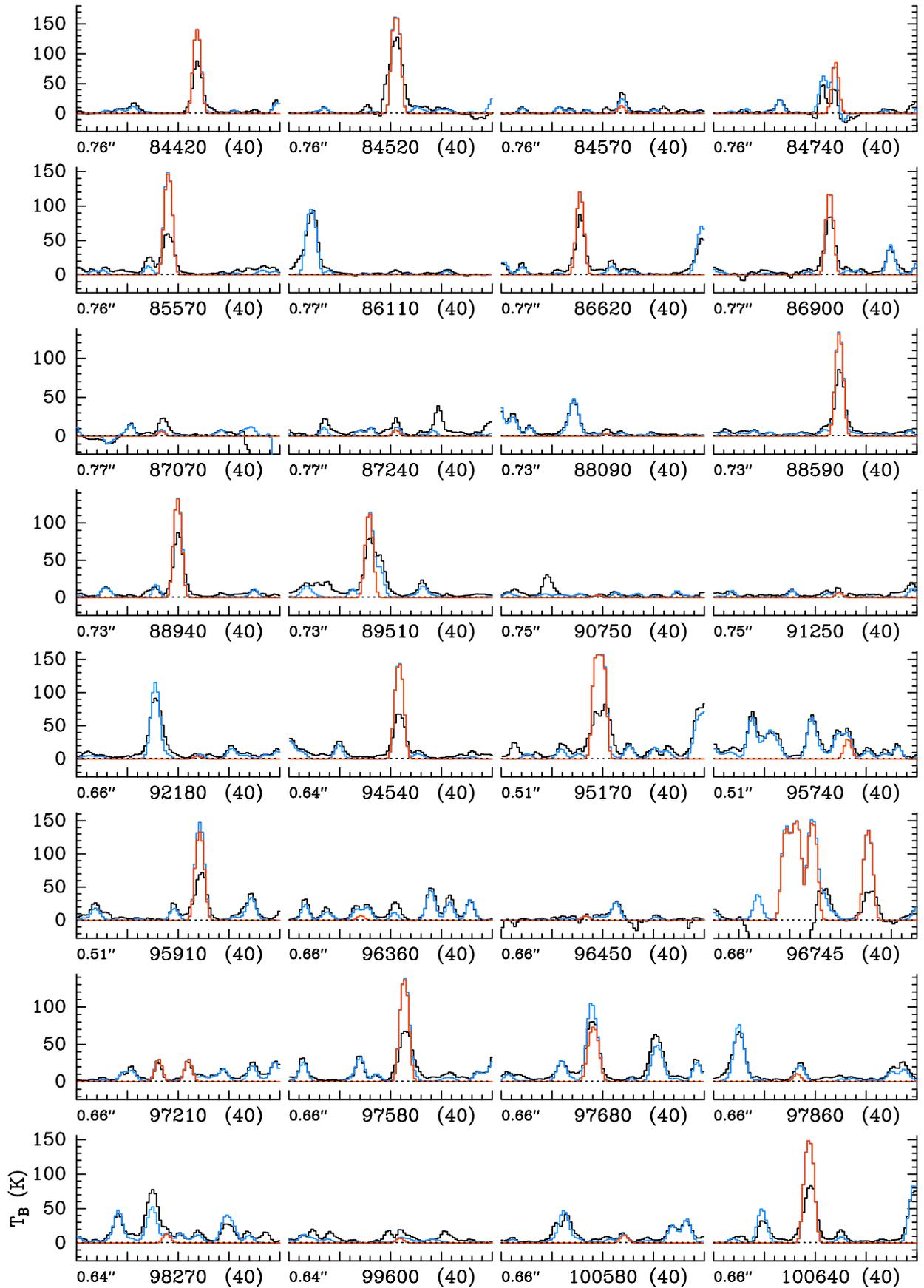}}}
\caption{Transitions of CH$_3$OH, $\varv_{\rm t} = 0$ covered by our ALMA 
survey. The best-fit LTE synthetic spectrum of CH$_3$OH, $\varv_{\rm t} = 0$ is 
displayed in red and overlaid on the observed spectrum of Sgr~B2(N1S), shown 
in black. The blue synthetic spectrum contains the contributions from all 
molecules identified in our survey so far, including from the species shown in red. 
The central frequency and width are indicated in MHz below each panel. The
angular resolution (HPBW) is also indicated. The $y$-axis is labeled in
brightness temperature units (K). The dotted line indicates the $3\sigma$ 
noise level.}
\label{f:spec_ch3oh_ve0_n1s}
\end{figure*}

\begin{figure*}[!ht]
\addtocounter{figure}{-1}
\centerline{\resizebox{0.88\hsize}{!}{\includegraphics[angle=0]{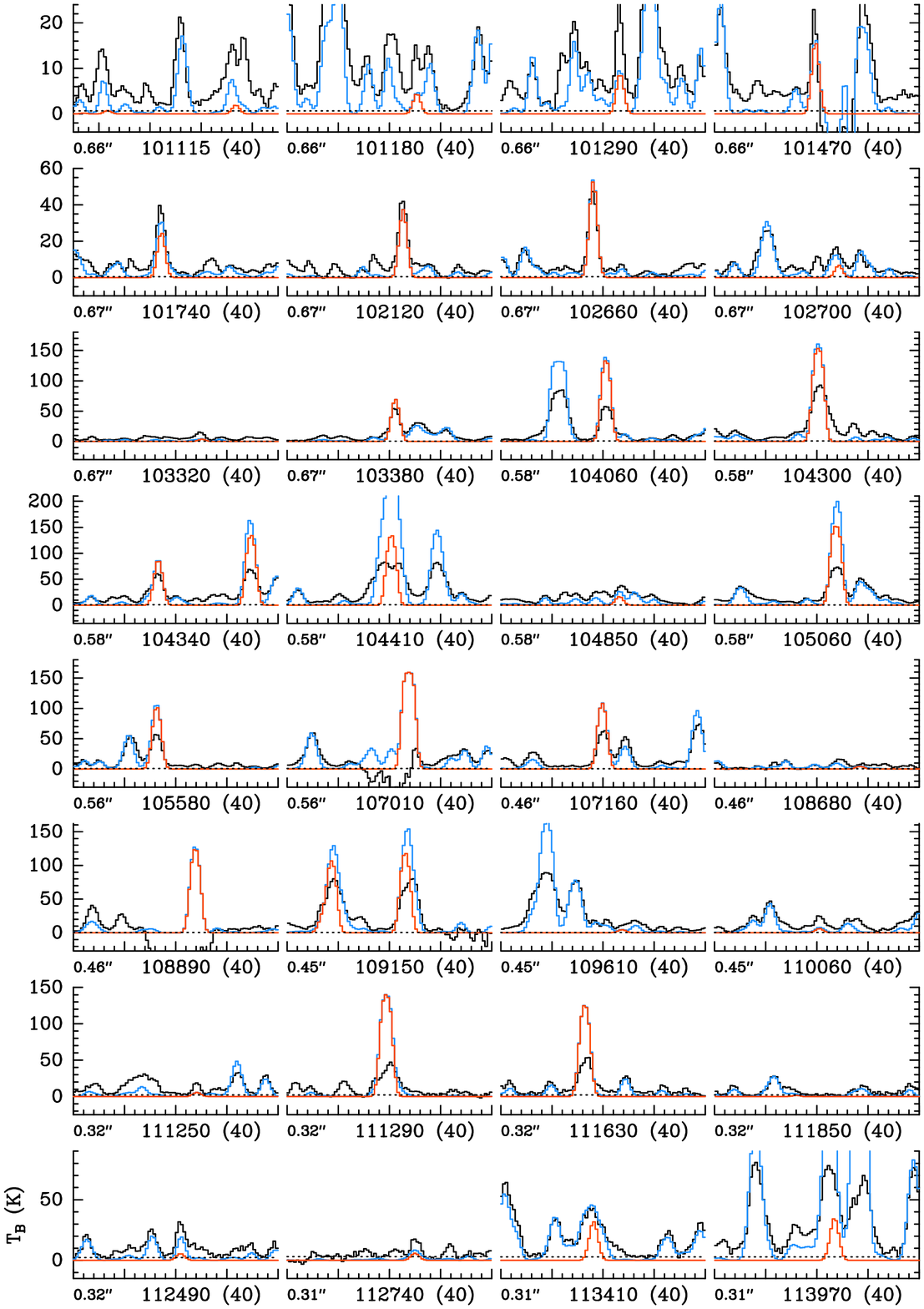}}}
\caption{continued.}
\end{figure*}

\begin{figure*}
\centerline{\resizebox{0.88\hsize}{!}{\includegraphics[angle=0]{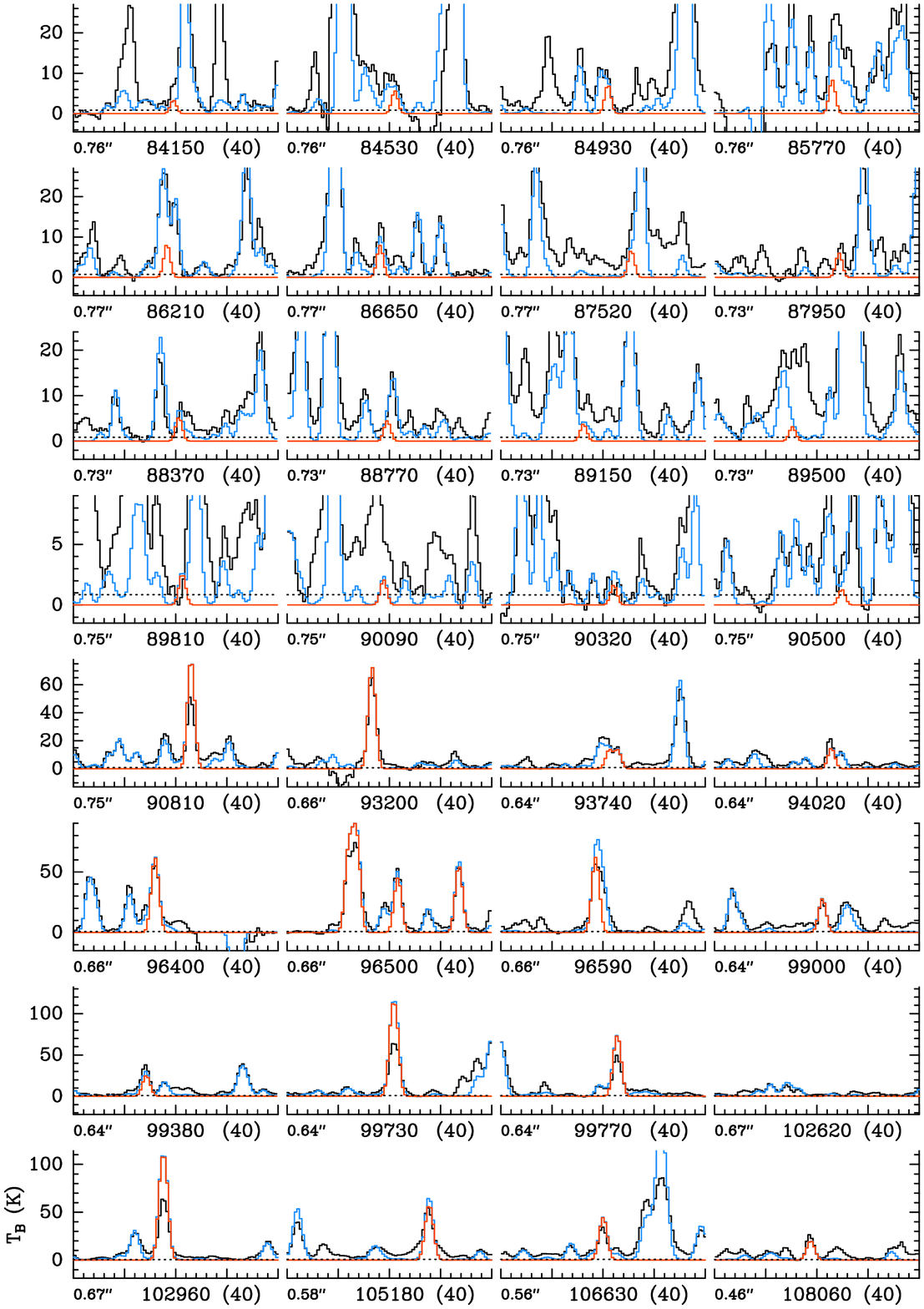}}}
\caption{Same as Fig.~\ref{f:spec_ch3oh_ve0_n1s}, but for CH$_3$OH, 
$\varv_{\rm t}=1$.}
\label{f:spec_ch3oh_ve1_n1s}
\end{figure*}

\begin{figure*}[!ht]
\addtocounter{figure}{-1}
\centerline{\resizebox{0.265\hsize}{!}{\includegraphics[angle=0]{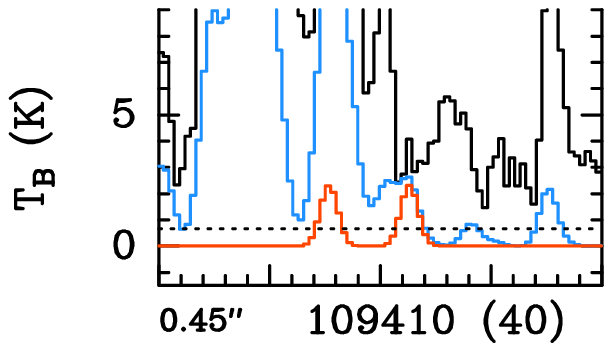}}}
\caption{continued.}
\end{figure*}

\begin{figure*}
\centerline{\resizebox{0.88\hsize}{!}{\includegraphics[angle=0]{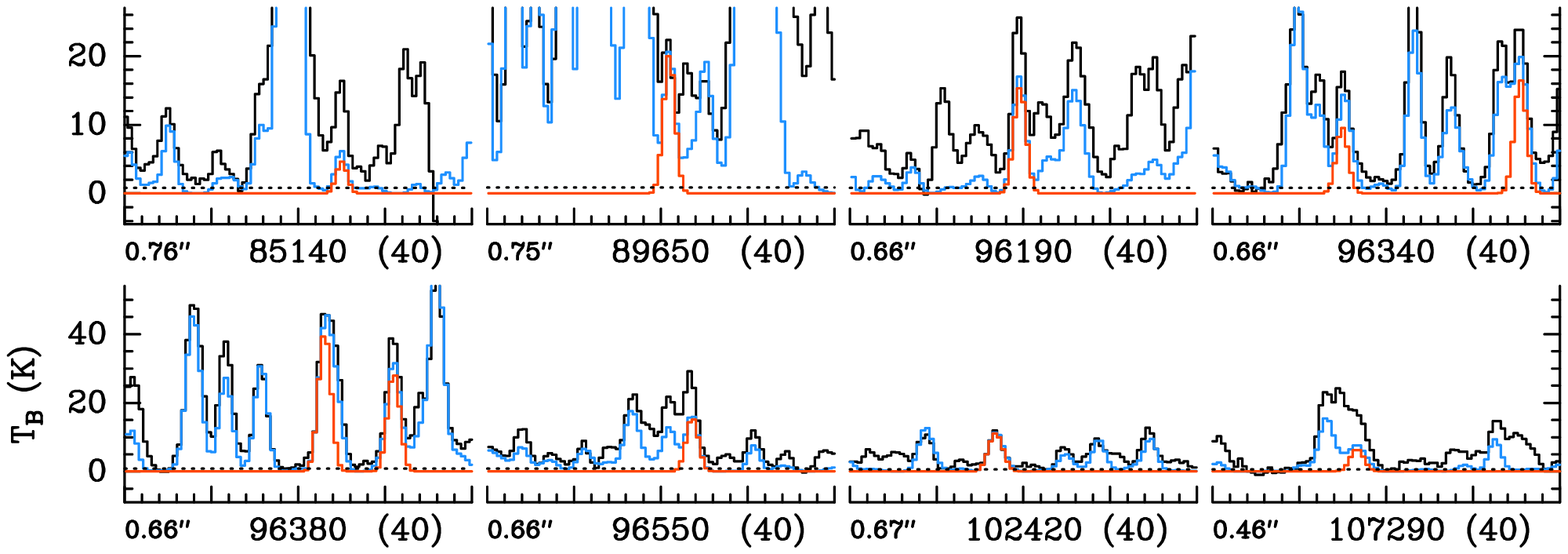}}}
\caption{Same as Fig.~\ref{f:spec_ch3oh_ve0_n1s}, but for CH$_3$OH, 
$\varv_{\rm t}=2$.}
\label{f:spec_ch3oh_ve2_n1s}
\end{figure*}

\begin{figure*}
\centerline{\resizebox{0.67\hsize}{!}{\includegraphics[angle=0]{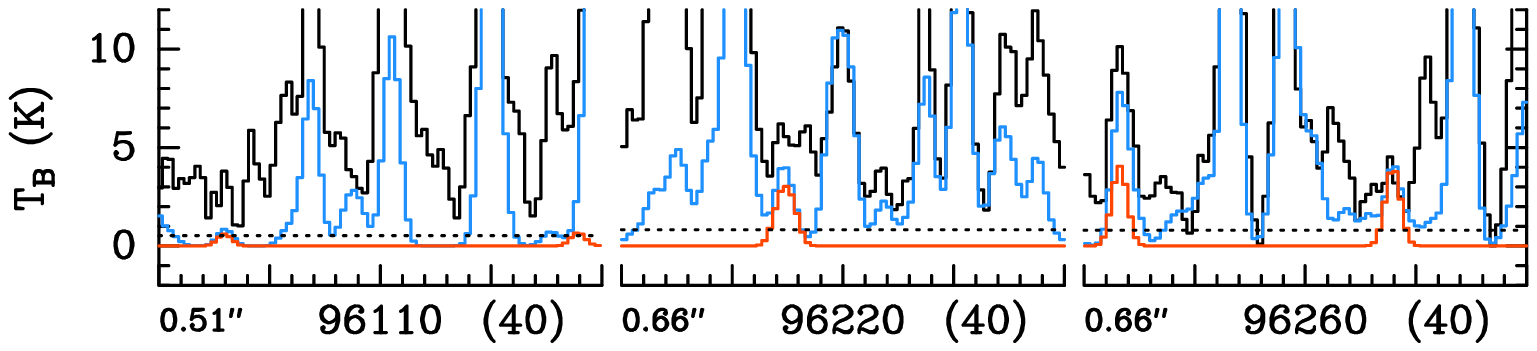}}}
\caption{Same as Fig.~\ref{f:spec_ch3oh_ve0_n1s}, but for CH$_3$OH, 
$\varv_{\rm t}=3$.}
\label{f:spec_ch3oh_ve3_n1s}
\end{figure*}

\begin{figure*}
\centerline{\resizebox{0.88\hsize}{!}{\includegraphics[angle=0]{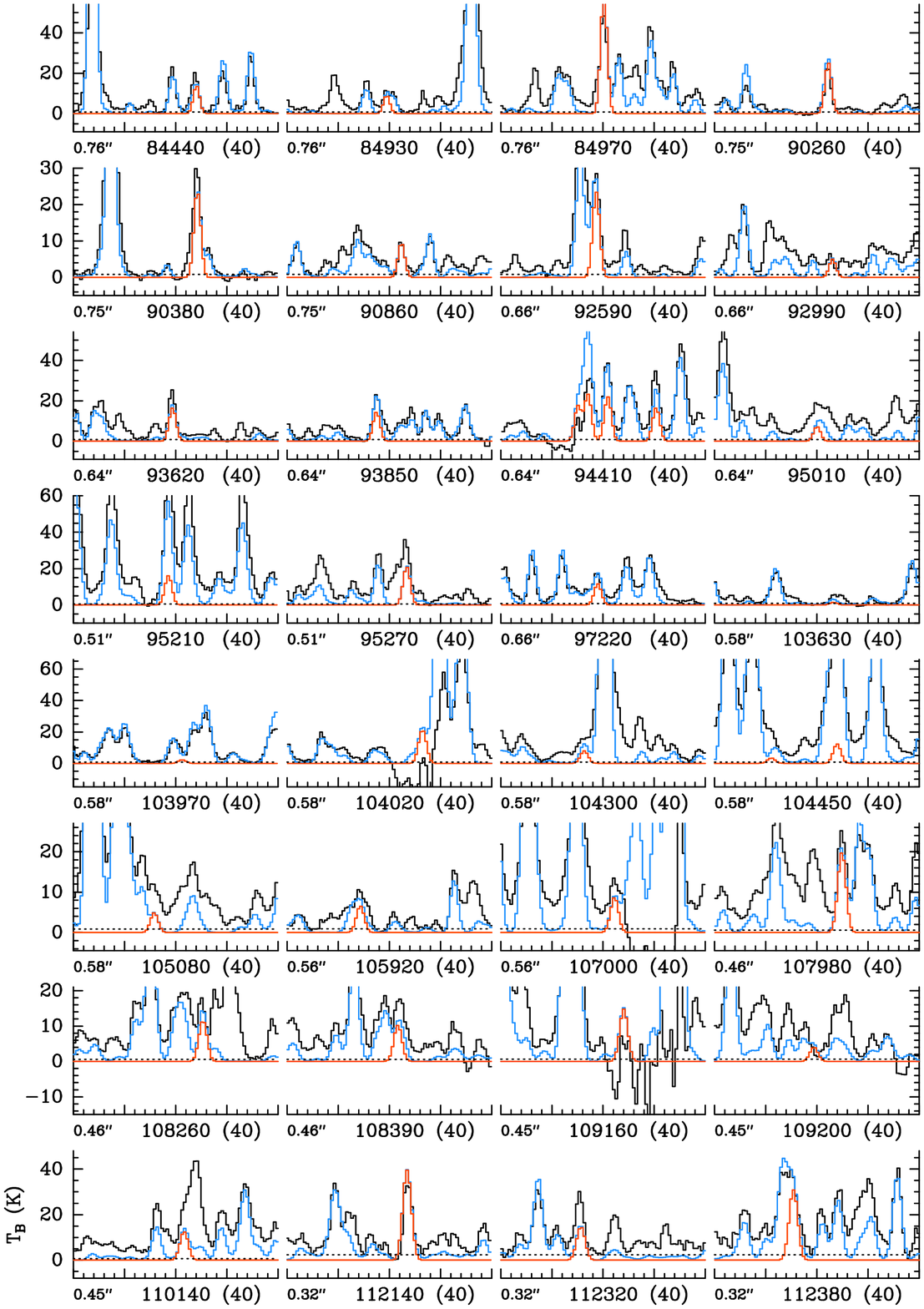}}}
\caption{Same as Fig.~\ref{f:spec_ch3oh_ve0_n1s}, but for $^{13}$CH$_3$OH, 
$\varv_{\rm t}=0$.}
\label{f:spec_ch3oh_13c_ve0_n1s}
\end{figure*}

\begin{figure*}[!ht]
\addtocounter{figure}{-1}
\centerline{\resizebox{0.265\hsize}{!}{\includegraphics[angle=0]{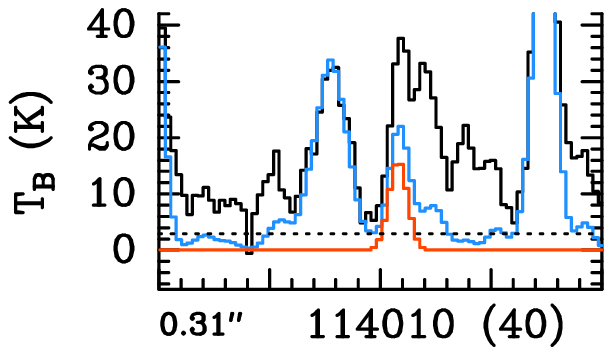}}}
\caption{continued.}
\end{figure*}

\begin{figure*}
\centerline{\resizebox{0.88\hsize}{!}{\includegraphics[angle=0]{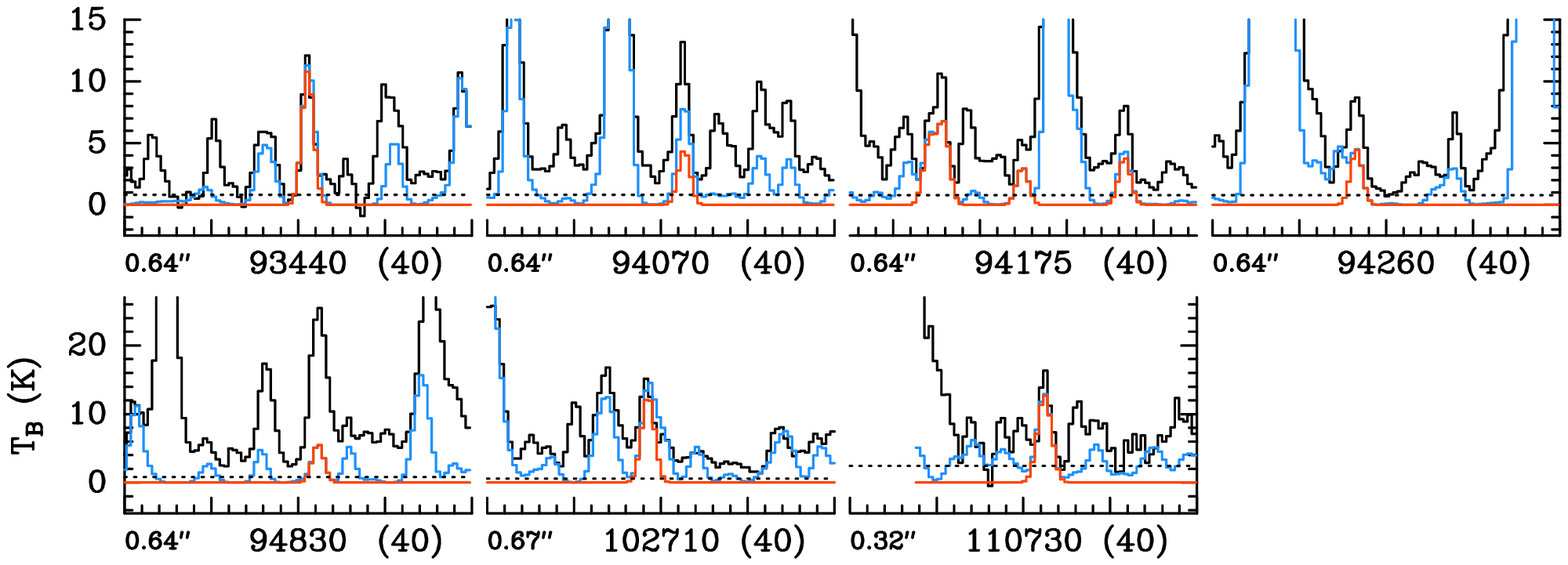}}}
\caption{Same as Fig.~\ref{f:spec_ch3oh_ve0_n1s}, but for $^{13}$CH$_3$OH, 
$\varv_{\rm t}=1$.}
\label{f:spec_ch3oh_13c_ve1_n1s}
\end{figure*}

\begin{figure*}
\centerline{\resizebox{0.88\hsize}{!}{\includegraphics[angle=0]{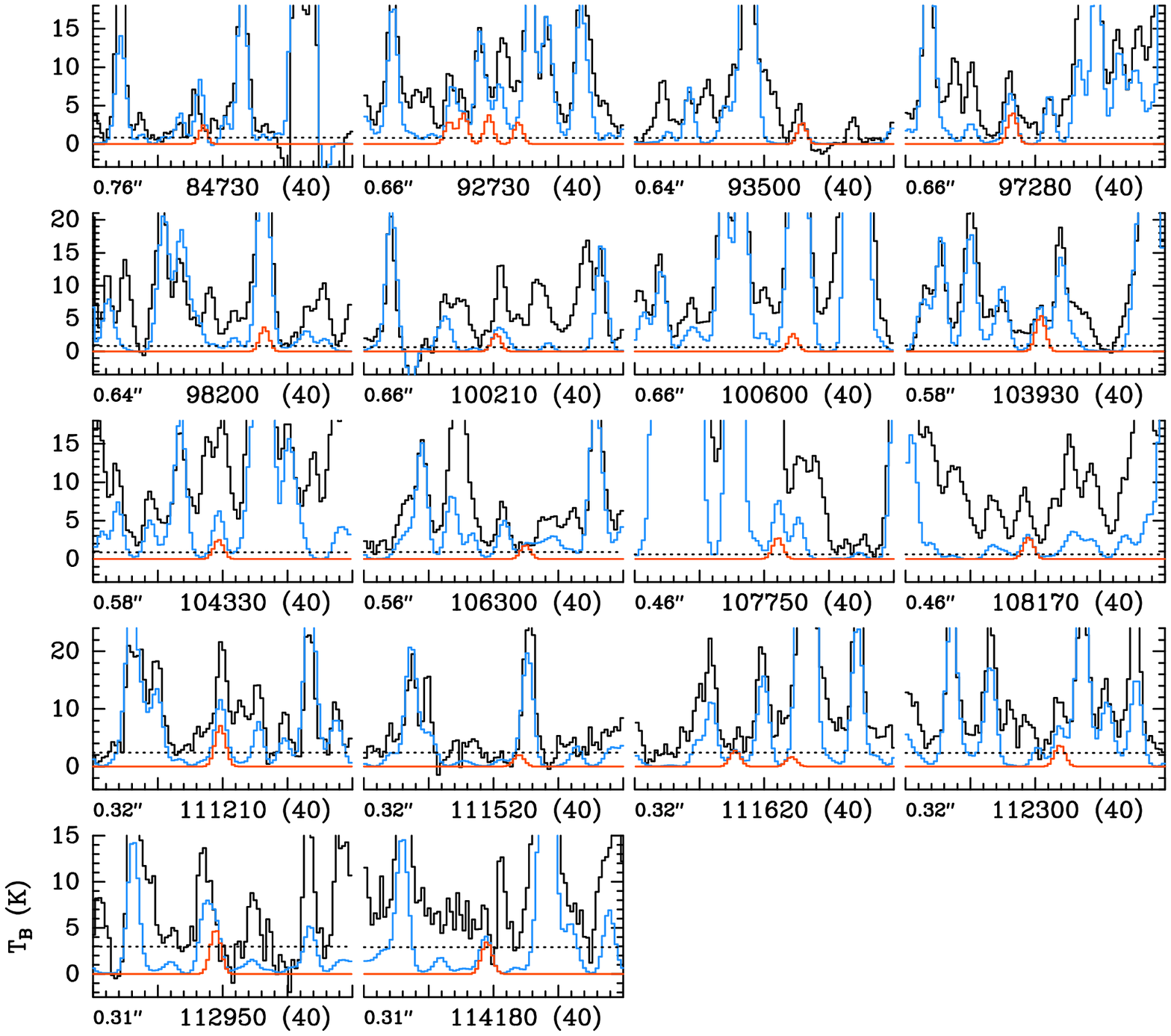}}}
\caption{Same as Fig.~\ref{f:spec_ch3oh_ve0_n1s}, but for CH$_3$$^{18}$OH, 
$\varv_{\rm t}=0$.}
\label{f:spec_ch3oh_18o_ve0_n1s}
\end{figure*}

\begin{figure*}
\centerline{\resizebox{0.88\hsize}{!}{\includegraphics[angle=0]{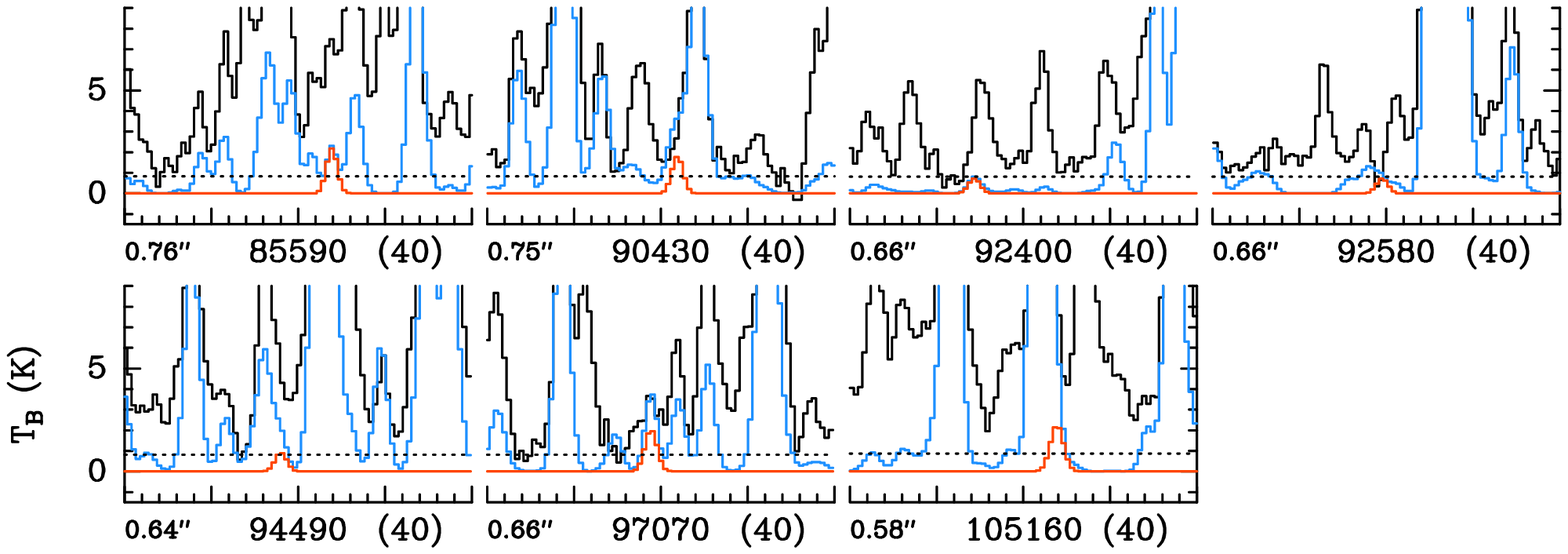}}}
\caption{Same as Fig.~\ref{f:spec_ch3oh_ve0_n1s}, but for CH$_3$$^{18}$OH, 
$\varv_{\rm t}=1$.}
\label{f:spec_ch3oh_18o_ve1_n1s}
\end{figure*}

\begin{figure*}
\centerline{\resizebox{0.88\hsize}{!}{\includegraphics[angle=0]{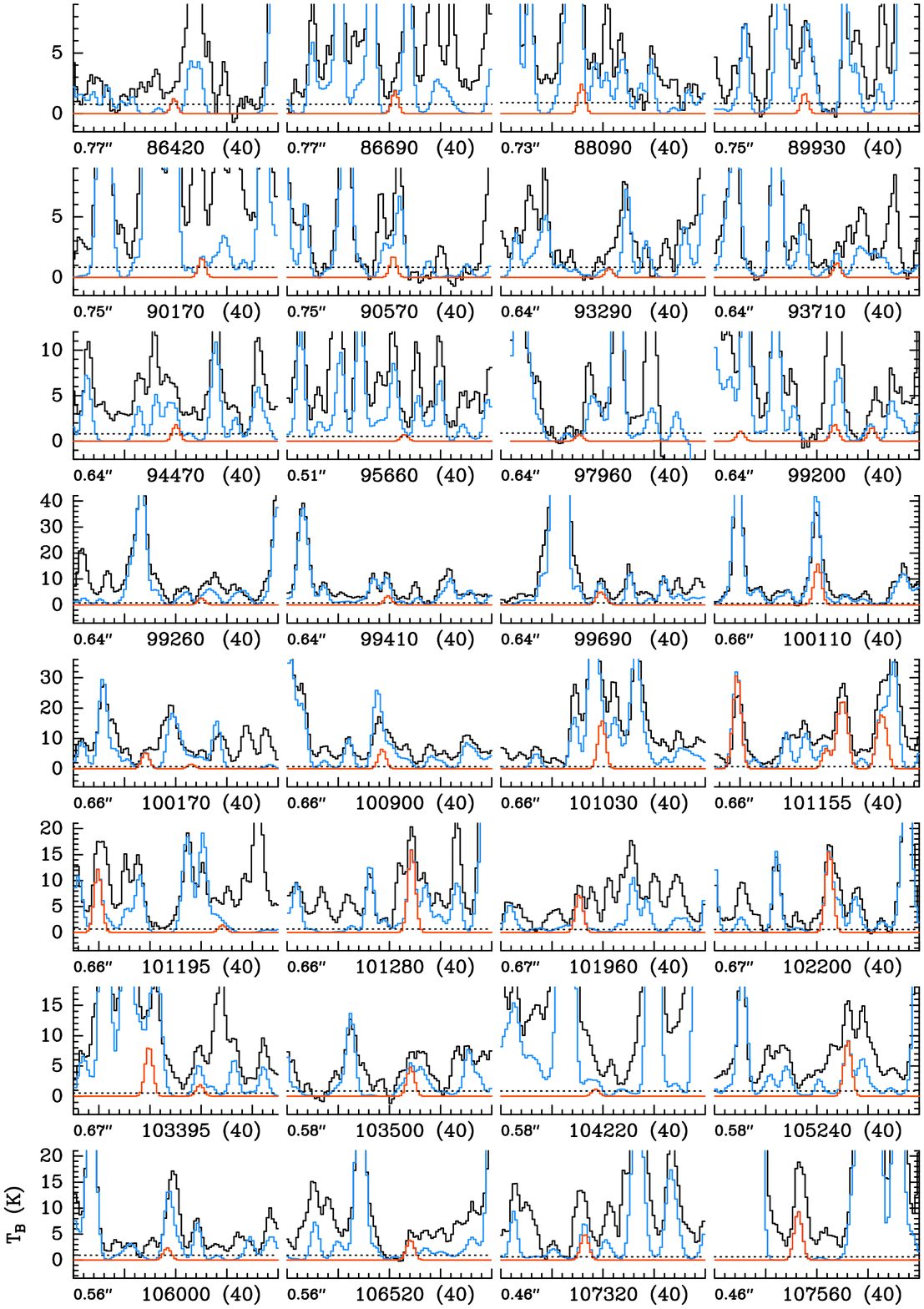}}}
\caption{Same as Fig.~\ref{f:spec_ch3oh_ve0_n1s}, but for CH$_3$SH, 
$\varv_{\rm t}=0$.}
\label{f:spec_ch3sh_ve0_n1s}
\end{figure*}

\begin{figure*}[!ht]
\addtocounter{figure}{-1}
\centerline{\resizebox{0.88\hsize}{!}{\includegraphics[angle=0]{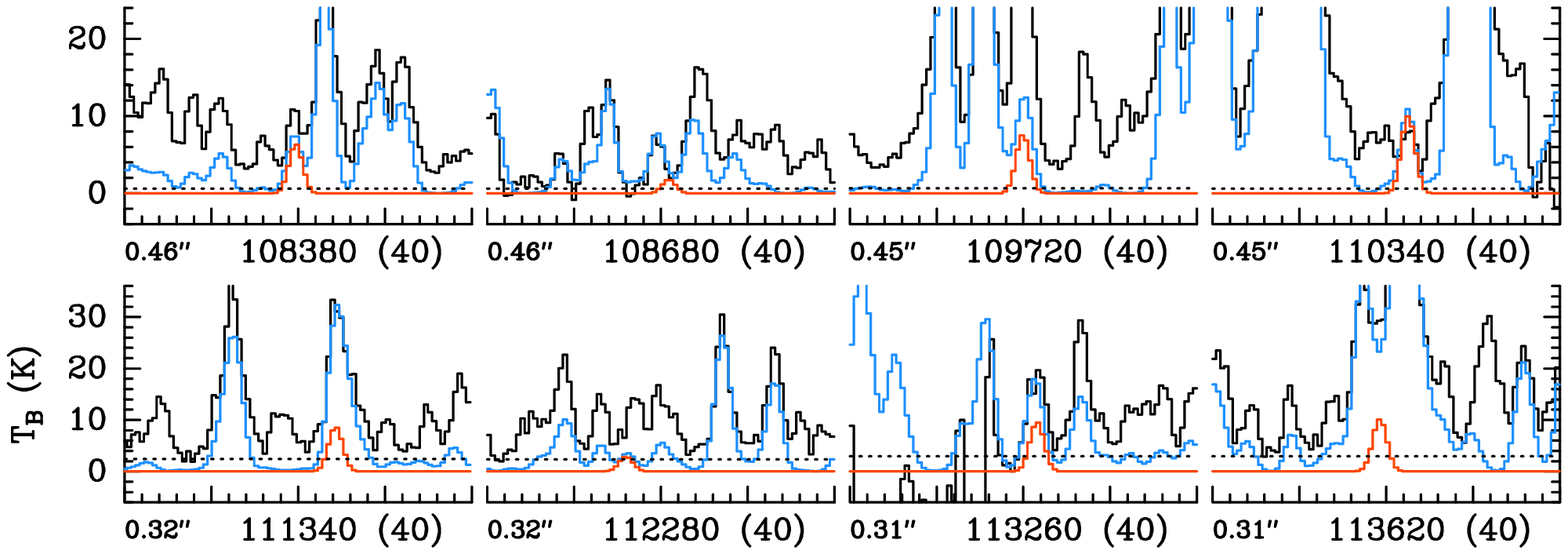}}}
\caption{continued.}
\end{figure*}

\begin{figure*}
\centerline{\resizebox{0.88\hsize}{!}{\includegraphics[angle=0]{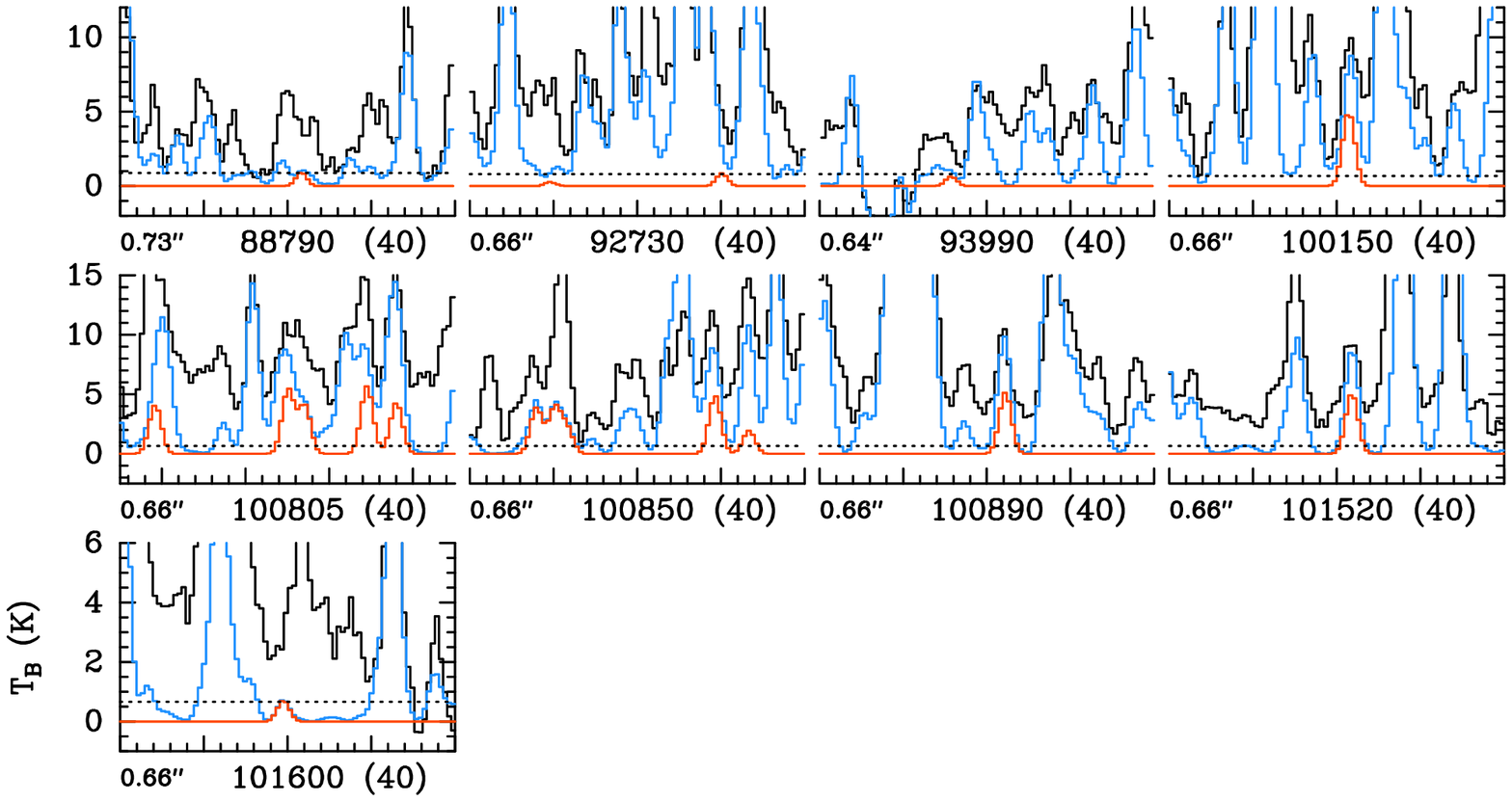}}}
\caption{Same as Fig.~\ref{f:spec_ch3oh_ve0_n1s}, but for CH$_3$SH, 
$\varv_{\rm t}=1$.}
\label{f:spec_ch3sh_ve1_n1s}
\end{figure*}

\begin{figure*}
\centerline{\resizebox{0.88\hsize}{!}{\includegraphics[angle=0]{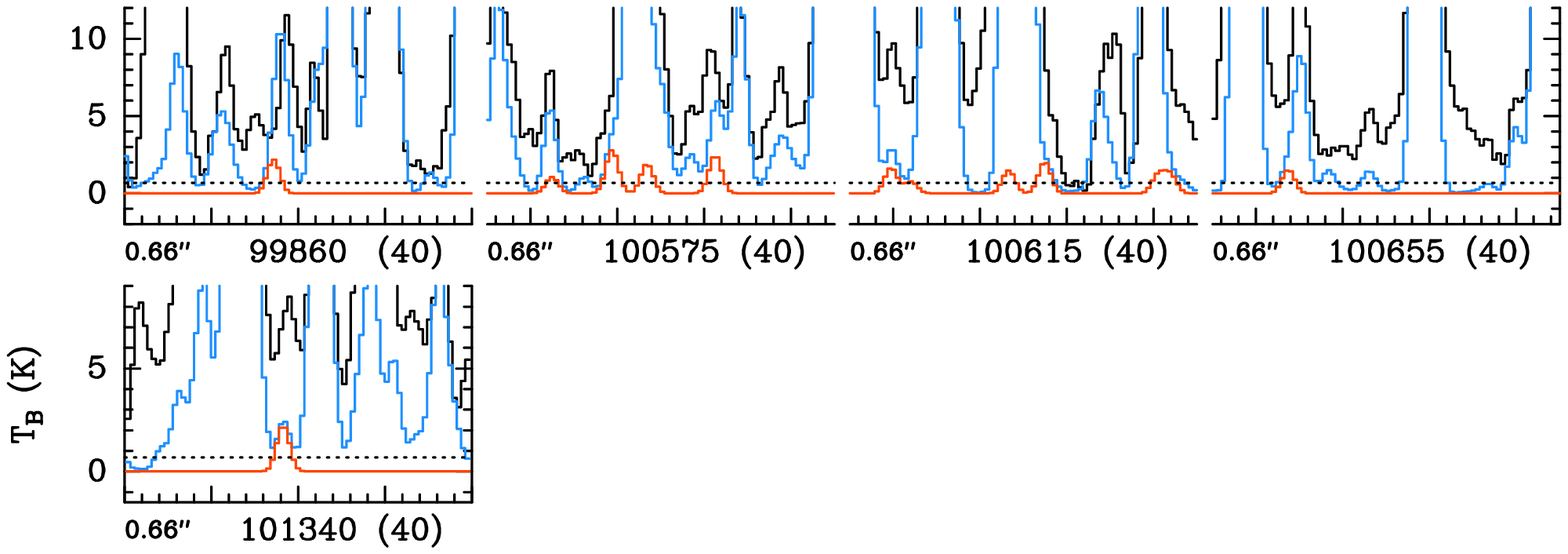}}}
\caption{Same as Fig.~\ref{f:spec_ch3oh_ve0_n1s}, but for CH$_3$SH, 
$\varv_{\rm t}=2$.}
\label{f:spec_ch3sh_ve2_n1s}
\end{figure*}

\section{Complementary figures: Population diagrams}
\label{a:popdiag}

\begin{figure}[!ht]
\centerline{\resizebox{0.95\hsize}{!}{\includegraphics[angle=0]{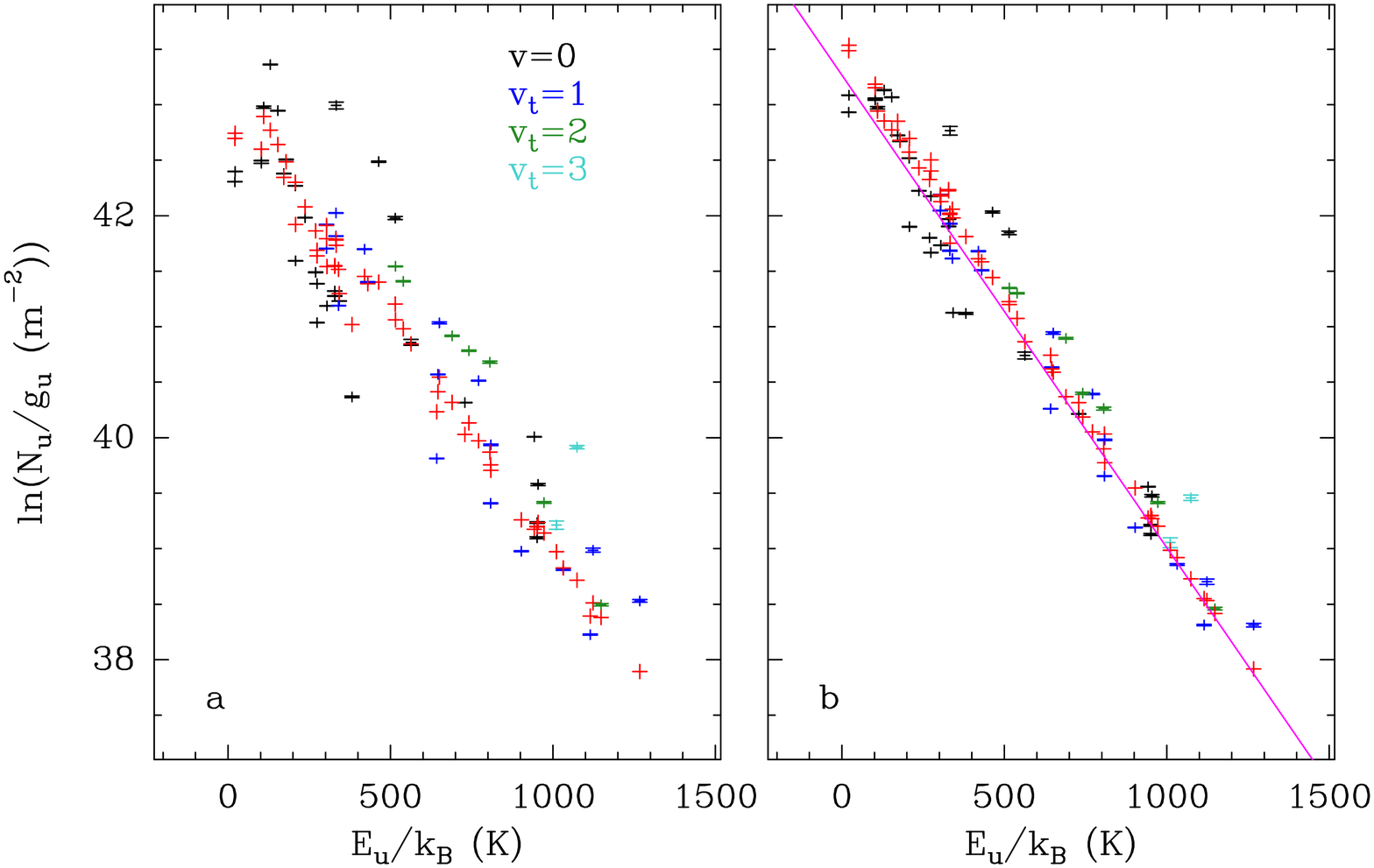}}}
\caption{Population diagram of CH$_3$OH toward Sgr~B2(N1S). The 
observed data points are shown in various colors (but not red), as indicated in 
the upper right corner of panel \textbf{a}, and the synthetic populations are 
shown in red. No correction is applied in panel \textbf{a}. 
In panel \textbf{b}, the optical depth correction has been applied to both the 
observed and synthetic populations, and the contamination by all other 
species included in the full model has been removed from the observed 
data points. The purple line is a linear fit to the observed populations (in 
linear-logarithmic space).
}
\label{f:popdiag_ch3oh_n1s}
\end{figure}

\begin{figure}[!ht]
\centerline{\resizebox{0.95\hsize}{!}{\includegraphics[angle=0]{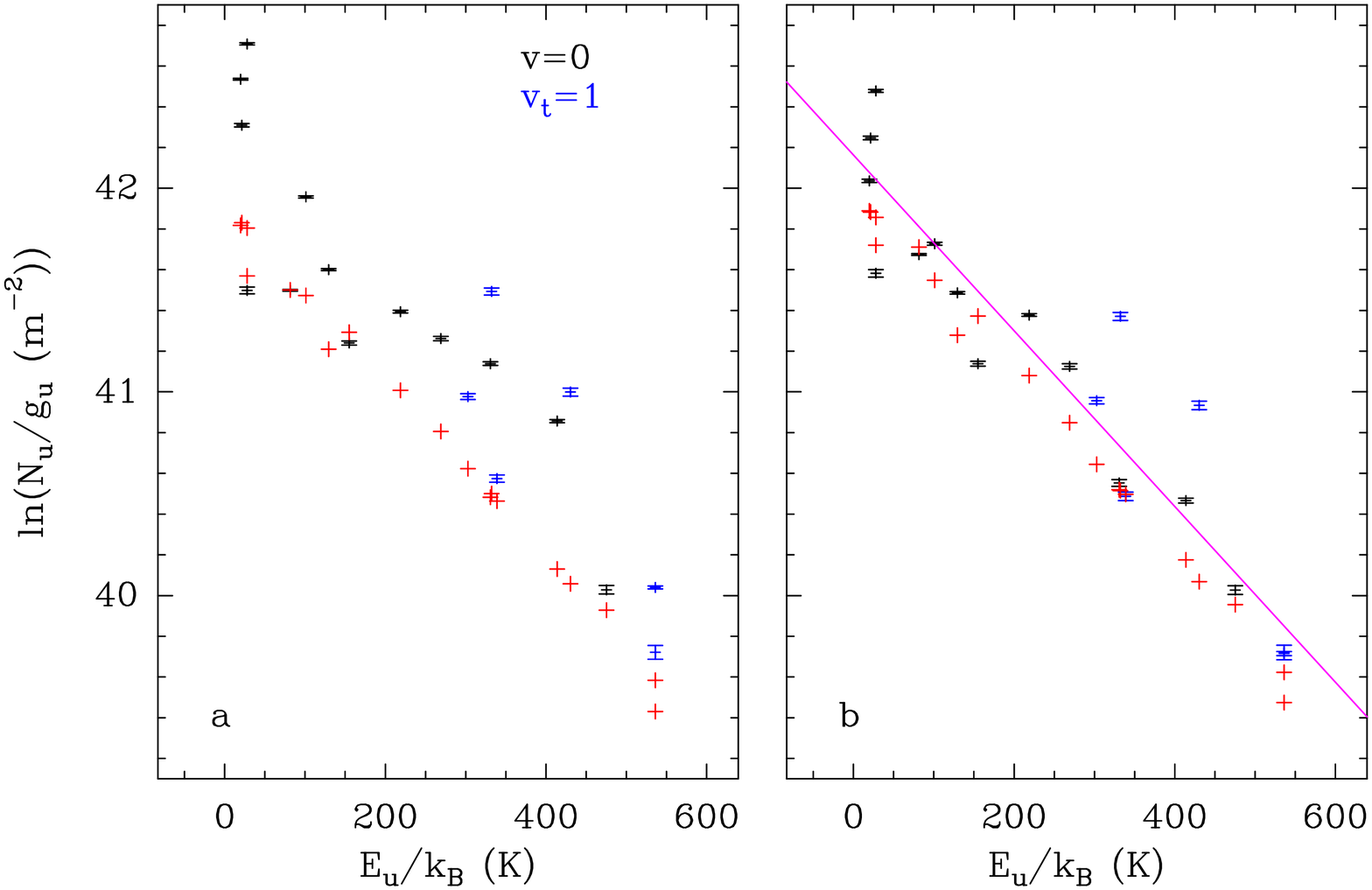}}}
\caption{Same as Fig.~\ref{f:popdiag_ch3oh_n1s}, but for $^{13}$CH$_3$OH.}
\label{f:popdiag_ch3oh_13c_n1s}
\end{figure}

\begin{figure}[!ht]
\centerline{\resizebox{0.95\hsize}{!}{\includegraphics[angle=0]{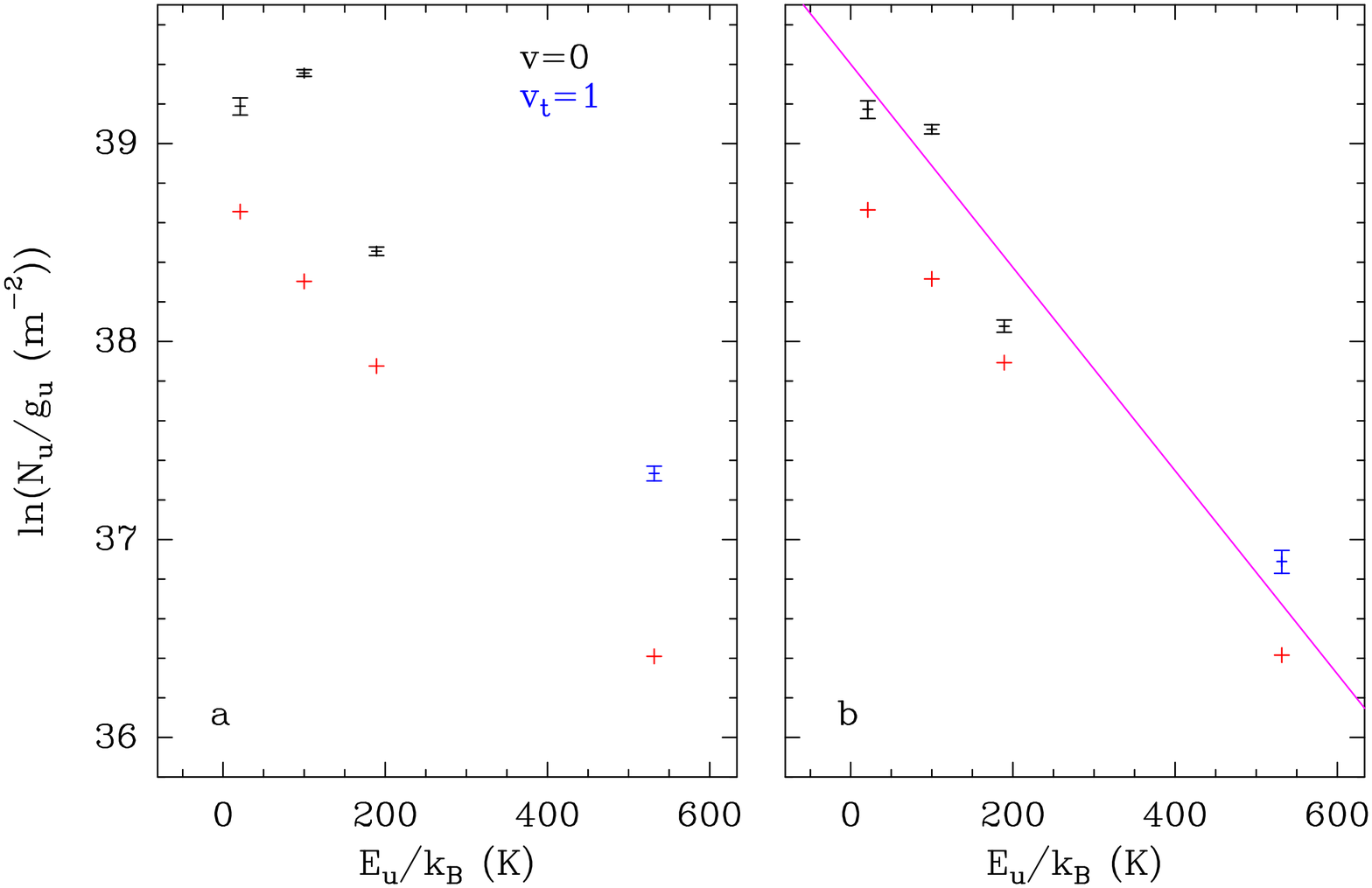}}}
\caption{Same as Fig.~\ref{f:popdiag_ch3oh_n1s}, but for CH$_3$$^{18}$OH.}
\label{f:popdiag_ch3oh_18o_n1s}
\end{figure}

\begin{figure}[!ht]
\centerline{\resizebox{0.95\hsize}{!}{\includegraphics[angle=0]{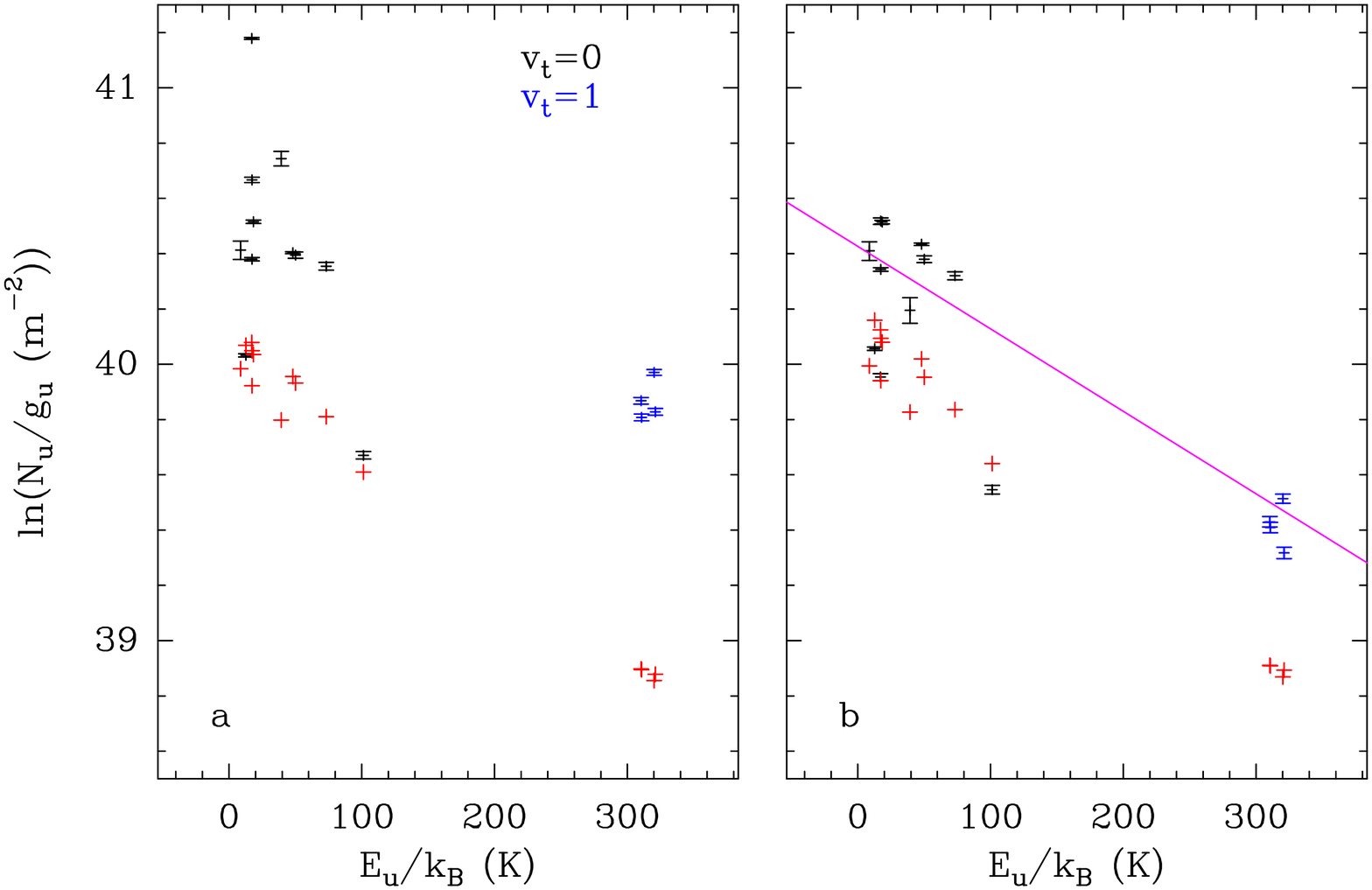}}}
\caption{Same as Fig.~\ref{f:popdiag_ch3oh_n1s}, but for CH$_3$SH.}
\label{f:popdiag_ch3sh_n1s}
\end{figure}

Figures~\ref{f:popdiag_ch3oh_n1s}--\ref{f:popdiag_ch3sh_n1s} show population
diagrams of methanol, its isotopologs, and methanethiol derived from
the ReMoCA survey toward  Sgr~B2(N1S).

\end{appendix}

\end{document}